%% file: DrawDown.tex
\documentclass[letterpaper]{article}
\usepackage{amsfonts,amsmath,amsthm,amssymb,mathtools,color}
\usepackage{mathrsfs}
\usepackage{graphicx}
\usepackage{geometry,appendix,adjustbox,float,array}
\usepackage{caption,subcaption}
\graphicspath{{figs/}}

\usepackage{plain}

\usepackage{dsfont}
\usepackage[round]{natbib}

\providecommand{\U}[1]{\protect\rule{.1in}{.1in}}
\numberwithin{equation}{section} 

\newcommand{\E}{\mathbb{E}\thinspace }

\newcommand{\F}{\mathbb{F}}

\begin{document}

\begin{plain}
\input{DrawDown_TeX.tex}
\end{plain}

\end{document}

%% file: DrawDown_TeX.tex
\magnification = 1100
\baselineskip = 15 pt

\def \noi {\noindent}

\def \F {{\cal F}}
\def \P {{\bf P}}
\def \E {{\bf E}}

\def \pv {\pi^*}
\def \la {\lambda}
\def \al {\alpha}
\def \be {\beta}
\def \del {\delta}
\def \gam {\gamma}

\def \R {{\bf R}}
\def \D {{\bf D}}
\def \C {{\cal C}}

\def \mems {\text{minimum expected maximum shortfall}}
\def \pr {\text{minimum probability of ruin}}

\def\sqr#1#2{{\vcenter{\vbox{\hrule height.#2pt\hbox{\vrule width.#2pt height#1pt \kern#1pt\vrule width.#2pt}\hrule height.#2pt}}}}

\def \square{\hfill\mathchoice\sqr56\sqr56\sqr{4.1}5\sqr{3.5}5}

\def \qed {$\square$ \medskip}

\def \sect#1{\bigskip \noindent {\bf  #1} \medskip}
\def \subsect#1{\medskip \noindent{\it #1} \medskip}
\def \th#1#2{\medskip \noindent {\bf  Theorem #1.}   \it #2 \rm \medskip}
\def \prop#1#2{\medskip \noindent {\bf  Proposition #1.}   \it #2 \rm \medskip}
\def \cor#1#2{\medskip \noindent {\bf  Corollary #1.}   \it #2 \rm \medskip}
\def \pf {\noindent  {\it Proof}.\quad }
\def \lem#1#2{\medskip \noindent {\bf  Lemma #1.}   \it #2 \rm \medskip}
\def \ex#1{\medskip \noindent {\bf  Example #1.}}
\def \rem#1 {\medskip \noi {\bf Remark #1.  }}

\centerline{\bf  Minimizing the Probability of Lifetime Drawdown under Constant Consumption} \bigskip \bigskip

\noi Bahman Angoshtari

Email: bango@umich.edu

\medskip

\noi Erhan Bayraktar

Email: erhan@umich.edu

\medskip

\noi Virginia R. Young

Email: vryoung@umich.edu

\bigskip

\noi 530 Church Street

\noi Department of Mathematics, University of Michigan

\noi Ann Arbor, Michigan, 48109

\bigskip

\centerline{Version: 17 May 2016}

\bigskip  \bigskip

\noindent{\bf  Abstract:}  We assume that an individual invests in a financial market with one riskless and one risky asset, with the latter's price following geometric Brownian motion as in the Black-Scholes model.  Under a constant rate of consumption, we find the optimal investment strategy for the individual who wishes to minimize the probability that her wealth drops below some fixed proportion of her maximum wealth to date, the so-called probability of {\it lifetime drawdown}.  If maximum wealth is less than a particular value, $m^*$, then the individual optimally invests in such a way that maximum wealth never increases above its current value.  By contrast, if maximum wealth is greater than $m^*$ but less than the safe level, then the individual optimally allows the maximum to increase to the safe level.

\medskip


\medskip


\medskip

\noi{\bf Journal of Economic Literature Classification:} G11 (primary); D14, D81 (secondary).

\sect{1. Introduction}

In 2008, many investors lost 30\% to 50\% of their wealth when the housing and stock markets greatly declined.  One-on-one conversations and news broadcasts continually focused on the regret that individuals felt as a result of those personal losses.  As a result, we were motivated to study the problem of minimizing the probability of so-called {\it lifetime drawdown}, that is, the probability that wealth drops below a given fraction of maximum wealth before death.  In most other research involving drawdown, wealth is constrained not to experience drawdown; see Grossman and Zhou (1993) and Cvitani\'c and Karatzas (1995) for early references, and see Kardaras et al.\ (2014) for a recent reference.  However, if the individual is consuming at a constant rate from her investment account, then one cannot prevent drawdown, so minimizing the probability of lifetime drawdown is a reasonable, objective goal.

In related research, Angoshtari et al.\ (2015) find the optimal investment strategy to minimize the probability of drawdown under {\it general} consumption with an infinite horizon.  Chen et al.\ (2015) is mostly closely related to the problem considered in this paper.  They minimize the probability of lifetime drawdown under two market assumptions; in the first, they consider two correlated risky assets with no consumption, and in the second, they consider a Black-Scholes market with consumption proportional to wealth.  In the latter case, they find that the optimal investment strategy does not allow maximum wealth to increase above its current level.  By contrast, we assume that the individual consumes at a {\it constant} rate.

The rest of the paper is organized as follows.  In Section 2, we introduce the financial market and define the problem of minimizing the probability of lifetime drawdown.  In Section 3, we prove a verification theorem for this minimum probability.  In Sections 4 and 5, we consider various cases for the values of wealth of the individual and the parameters of the model and solve the problem in those cases.  If maximum wealth is less than a particular value, $m^*$, then the individual optimally invests in such a way that maximum wealth never increases above its current value.  On the other hand, if maximum wealth is greater than $m^*$ but less than the safe level, then the individual optimally allows the maximum to increase to the safe level.  Section 6 concludes the paper.

\sect{2. Financial market and probability of lifetime drawdown}

In this section, we first present the financial ingredients that affect the individual's wealth, namely, consumption, a riskless asset, and a risky asset.  Then, we define the minimum probability of lifetime drawdown.  We assume that the individual invests in a riskless asset that earns interest at a constant rate $r > 0$.  Also, the individual invests in a risky asset whose price at time $t$, $S_t$, follows geometric Brownian motion with dynamics
$$
dS_t = S_t (\mu dt + \sigma dB_t),
$$
in which $\mu > r$, $\sigma > 0$, and $B$ is a standard Brownian motion with respect to a filtration of a probability space $(\Omega, {\cal F}, \P)$.

Let $W_t$ denote the wealth of the individual's investment account at time $t \ge 0$.  Let $\pi_t$ denote the dollar amount invested in the risky asset at time $t \ge 0$.  An investment policy $\Pi = \{ \pi_t \}_{t \ge 0}$ is {\it admissible} if it is an $\F$-progressively measurable process satisfying $\int_0^t \pi^2_s \, ds < \infty$ almost surely, for all $t \ge 0$.  It follows that the amount invested in the riskless asset at time $t \ge 0$ is $W_t - \pi_t$.  We assume that the individual consumes at a (net) constant rate $c > 0$.  Therefore, the wealth process follows
$$
dW_t = \left( rW_t + (\mu - r) \pi_t  - c \right) dt + \sigma \pi_t dB_t,
$$
and we suppose that initial wealth is non-negative; that is, $W_0 = w \ge 0$.

Define the maximum wealth $M_t$ at time $t$ by
$$
M_t = \max \left[ \sup_{0 \le s \le t} W_s, \; M_0 \right],
$$
\noindent in which we include $M_0 = m > 0$ (possibly different from $W_0 = w$) to allow the individual to have a financial past.  By {\it lifetime drawdown}, we mean that the individual's wealth reaches $\al \in [0, 1)$ times maximum wealth before she dies.  Define the corresponding hitting time by $\tau_\al := \inf\{t \ge 0: W_t \le \al M_t\}$.  Let $\tau_d$ denote the random time of death of the individual.  We assume that $\tau_d$ is exponentially distributed with parameter $\la$ (that is, with expected time of death equal to $1/\la$); this parameter is also known as the {\it hazard rate}.

\rem{2.1} {Moore and Young (2006) minimize the probability of ruin with varying hazard rate and show that by updating the hazard rate each year and treating it as a constant, the agent can quite closely obtain the minimal probability of ruin when the true hazard rate is Gompertz.  Specifically, at the beginning of each year, set $\la$ equal to the inverse of the agent's life expectancy at that time.  Compute the corresponding optimal investment strategy as given below, and apply that strategy for the year.  According to the work of Moore and Young (2006), this scheme results in a  probability of ruin close to the minimum probability of ruin.   Therefore, there is no significant loss of generality to assume that the hazard rate is constant and to revise its estimate each year.  Also, in the setting of an endowment fund of an organization, the assumption that the hazard rate for the organization is constant is not unreasonable.  \qed}

Denote the minimum probability of lifetime drawdown by $\phi(w, m)$, in which the arguments $w$ and $m$ indicate that one conditions on the individual possessing wealth $w$ at the current time, with maximum (past) wealth $m$.  Thus, $\phi$ is the minimum probability that $\tau_\al < \tau_d$, in which one minimizes with respect to admissible investment strategies $\pi$. Thus, $\phi$ is formally defined by
$$
\phi(w, m) = \inf_{\pi} \P^{w, m} \left(\tau_\al < \tau_d \right),
\eqno(2.1)
$$
for $w \le m$.  Here, $\P^{w, m}$ indicates the probability conditional on $W_0 = w$ and $M_0 = m$.  Below, we similarly write $\E^{w, m}$ for the corresponding conditional expectation.

Note that we may rewrite $\phi$ as
$$
\eqalign{
\phi(w, m) &= \inf_\pi \E^{w, m} \left[  \int_0^\infty  {\bf 1}_{\{ \tau_\al < t \}} \, \la \, e^{-\la t} \, dt \right]
= \inf_\pi \E^{w, m} \left[  \int_{\tau_\al}^\infty  {\bf 1}_{\{ \tau_\al < \infty \}} \, \la \, e^{-\la t} \, dt \right] \cr
&= \inf_\pi \E^{w, m} \left[  e^{-\la \tau_\al}  {\bf 1}_{\{ \tau_\al < \infty \}} \right] = \inf_\pi \E^{w, m} \left[  e^{-\la \tau_\al} \right].}
\eqno(2.2)
$$
This alternative representation will be useful in proving the verification theorem in the next section.

\rem{2.2} {If $\al = 0$, then the problem becomes one of minimizing the probability of lifetime ruin under constant consumption, as studied in Young (2004) and in Bayraktar and Young (2007), for example.}


\sect{3. Verification theorem}

In this section, we prove a verification theorem for the minimum probability of lifetime drawdown.  First, define the differential operator ${\cal L}^\beta$ for $\beta \in \R$ by
$$
{\cal L}^\beta f = (rw + (\mu - r) \beta - c) f_w + {1 \over 2} \sigma^2 \beta^2 f_{ww} - \la f,
$$
in which $f = f(w, m)$ is twice-differentiable with respect to its first variable.

Throughout the remainder of the paper, assume that $w > \al m$; otherwise, drawdown has occurred, and the game is over.  Also, note that if $w \ge c/r$, then drawdown is impossible.  Indeed, in that case, if the individual puts all her wealth $W_t$ in the riskless asset for $t \ge 0$ and consumes at the rate of $c$ from the investment earnings of $r W_t \ge c$, then wealth will steadily increase (or not decrease).  In other words, if $w \ge c/r$, then wealth never drops to $\al M_t$ almost surely, for all $t \ge 0$.  It follows that $\phi$ is identically $0$ when $w \ge c/r$.  Thus, we need only consider $\phi$ on the domain $\D := \{(w,m) \in ({\bf R}^+)^2: \al m \le w \le \min(m, c/r) \}$. 



\th{3.1} {Suppose $h: \D \rightarrow {\bf R}$ is a bounded, continuous function that satisfies the following conditions.
\item{$(i)$} $h(\cdot, m) \in \C^2\Big(\big(\al m, \min(m, c/r)\big)\Big)$ is non-increasing and convex,
\item{$(ii)$} $h(w, \cdot)$ is continuously differentiable, except possibly at finitely many values of $m\in[0,c/r]$, where it has (bounded) right- and left-derivatives,
\item{$(iii)$} $h_m(m, m) \ge 0$ if $m < c/r$ and $h_m(m, m)$ exists,
\item{$(iv)$} $h(\al m, m) = 1$,
\item{$(v)$} $h(c/r, m) = 0$ if $m \ge c/r$,
\item{$(vi)$} ${\cal L}^\beta h \ge 0$ for all $\beta \in {\bf R}$.}

\noi{\it Then, $h \le \phi$ on $\D$.}

\medskip

\pf Assume that $h$ satisfies the conditions specified in the statement of this theorem.  Let $W^\pi$ and $M^\pi$ denote the wealth and the maximum wealth, respectively, when the individual uses an admissible investment policy $\pi$.  Also, assume that the ordered pair of initial wealth and maximum wealth $(w, m)$ lie in $\D$.

Fix an admissible investment policy $\pi$.  Define $\tau_n = \inf \{t \ge 0:  \int_0^t \pi^2_s \, ds \ge n \}$, $\tau_{c/r} = \inf \{ t \ge 0: W_t^\pi = c/r \}$, and $\tau = \tau_\al \wedge \tau_n \wedge \tau_{c/r}$.  By applying It\^{o}'s formula to $e^{-\la t} h(w, m)$, we have
\begin{equation}\label{eq:Ito}\tag{3.1}
\eqalign{
e^{-\la \tau} h(W^\pi_{\tau}, M^\pi_{\tau}) &= h(w,m) + \int_{0}^{\tau} e^{-\la t} \, h_w(W^\pi_t, M^\pi_t) \, \sigma \, \pi_t \, dB_t  \cr
& \quad +\int_{0}^{\tau} e^{-\la t} \, {\cal L}^\pi h(W^\pi_t, M^\pi_t) \, dt  + \int_{0}^{\tau} e^{-\la t} \, h^-_m (W^\pi_t, M^\pi_t) \, dM^\pi_t.}
\end{equation}
Here, we used the fact that $M^\pi$ is continuous and $h_m$ exists almost everywhere ($h^-_m$ denotes the left derivative).  Also, since $M^\pi$ is non-decreasing, the first variation process associated with it is finite almost surely, and we conclude that the cross variation of $M^\pi$ and $W^\pi$ is zero almost surely.

It follows from the definition of $\tau_n$ that
$$
\E^{w, m} \left[\int_{0}^{\tau} e^{-\la t} \, h_w(W^\pi_t, M^\pi_t) \, \sigma \, \pi_t \, dB_t \right] = 0.
$$
Also, the second integral in \eqref{eq:Ito} is non-negative because of condition (vi) of the theorem.  Finally, the third integral  is non-negative almost surely because $dM_t$ is non-zero only when $M_t = W_t$ and $h_m(m,m) \ge 0$, almost everywhere, by condition (iii). Thus, we have
$$
\E^{w,m} [e^{-\la \tau} h(W^\pi_{\tau}, M^\pi_{\tau})] \ge h(w,m).
$$

Because $h$ is bounded by assumption, it follows from the dominated convergence theorem that
$$
\E^{w,m}[e^{-\la (\tau_\al \wedge \tau_{c/r})} h(W^\pi_{\tau_\al \wedge \tau_{c/r}}, M^\pi_{\tau_\al \wedge \tau_{c/r}})] \ge h(w,m).
$$
Since $W^\pi_{\tau_\al} = \al M^\pi_{\tau_\al}$ and $W^\pi_{\tau_{c/r}} = c/r$ when $(W^\pi_0, M^\pi_0) = (w, m) \in \D$, it follows from conditions (iv) and (v) of the theorem that
$$
h(w, m) \le \E^{w,m}[e^{-\la \tau_\al} {\bf 1}_{\{\tau_\al < \tau_{c/r}\}} ] = \E^{w,m}[e^{-\la \tau_\al} ].
\eqno(3.2)
$$
The equality in (3.2) follows from the fact that $\tau_\al = \infty$ if $\tau_{c/r} \le \tau_\al$.  By taking the infimum over admissible investment strategies, and by applying the representation of $\phi$ from (2.2), we obtain $h \le \phi$ on $\D$.  \qed

We use this verification theorem in the case embodied by the following corollary of Theorem 3.1.



\cor{3.2}{Suppose $h$ satisfies the conditions of Theorem $3.1$ in such a way that conditions $(iii)$ and $(vi)$ hold with equality, for some admissible strategy $\pi$ defined in feedback form by $\pi_t = \pi(W_t, M_t),$ in which we slightly abuse notation.  Then, $h = \phi$ on $\D$, and $\pi$ is an optimal investment strategy.}

\pf In the proof of Theorem 3.1, if we have equality in conditions (iii) and (vi), then we can conclude that $h = \phi$ on $\D$.  \qed

In the next two sections, we use this verification theorem and its corollary to determine the minimum probability of drawdown $\phi$.

\sect{4.  Minimum probability of drawdown when $m \ge c/r$}

In this section, we consider the case for which $m \ge c/r$; recall that $w \le c/r$.  Define an investment strategy $\pi$ as a feedback control, with a slight abuse of notation, as follows.
$$
\pi_t = \pi(W^\pi_t) = {\mu - r \over \sigma^2} \, {1 \over \gam - 1} \, \left( {c \over r} -  W^\pi_t \right),
\eqno(4.1)
$$
in which $\gam$ is defined by
$$
\gam = {1 \over 2r} \left[ (r + \la + \del) + \sqrt{(r + \la + \del)^2 - 4r \la} \right] > 1,
$$
with
$$
\del = {1 \over 2} \left( {\mu - r \over \sigma} \right)^2.
$$
Recall that $W^\pi$ and $M^\pi$ denote the wealth and maximum wealth, respectively, under the investment strategy $\pi$. 

One can show that $W^\pi$ follows the process
$$
dW^\pi_t = \left( {c \over r} - W^\pi_t \right) \left\{  \left( {2 \del \over \gam - 1} - r \right) dt + {\mu - r \over \sigma} \, {1 \over \gam - 1} \, dB_t \right\}.
$$
Note that if $W_0 = w < c/r$, we have $W^\pi_t < c/r$ almost surely, for all $t \ge 0$, under this investment strategy.  Thus, $M^\pi_t = m$ almost surely, for all $t \ge 0$.  From Young (2004), we know that the probability of drawdown under this strategy is given by
$$
h(w, m) = \left( {c/r - w \over c/r - \al m} \right)^\gam, \quad \al m \le w \le c/r.
\eqno(4.2)
$$
In the next theorem, we show that $h$ is the {\it minimum} probability of drawdown.

\th{4.1} {When $m \ge c/r,$ the minimum probability of drawdown $\phi$ on $\D = \{ (w, m) \in (\R^+)^2: \al m \le w \le c/r \}$ is given by the expression in $(4.2)$.  An optimal investment strategy $\pi$ is given in feedback form by $(4.1)$.}

\pf  It is straightforward to show that $h$ in (4.2) satisfies conditions (i), (ii), (iv), (v), and (vi) of Theorem 3.1, the last with equality when $\beta = \pi(w)$ in (4.1).  Condition (iii) is moot because $m \ge c/r$.  Thus, by Corollary 3.2, $\phi = h$ in (4.2) on $\D$, with an optimal investment strategy $\pi$ given in (4.1).  \qed

Theorem 4.1 tells us that when wealth is less than the so-called {\it safe level} $c/r$, and when that safe level is less than the maximum wealth $m$, in order to minimize the probability of drawdown, the individual's wealth cannot reach the safe level and, thereby, cannot reach a new maximum.  The individual effectively treats her drawdown level, $\al m$, as a constant ruin level, and the results of Young (2004) apply.

It follows from the investment strategy given in (4.1), that as wealth increases towards $c/r$, the amount invested in the risky asset decreases to zero. This makes sense because as the individual becomes wealthier, she does not need to take on as much risk to achieve her fixed consumption rate of $c$.

\setcounter{section}{5}
\sect{5. Minimum probability of drawdown when $0 < m < c/r$}

In the previous section, we showed that it is optimal for $M_t = m$ almost surely, for all $t \ge 0$, when $0 < w < c/r \le m$.  In this section, we show that if $m\in(0,c/r)$ is large enough, then it is optimal to allow $M$ to increase above $m$. In particular, we show that there exists a \emph{critical high-water mark} $m^*\in(0,c/r)$ with the following properties.

\item{$(i)$} If $m\in(m^*,c/r)$, then the optimal investment strategy allows for $M$ to increase above $m$; and
\item{$(ii)$} If $m\in(0,m^*]$, then the optimal investment strategy does not allow $M$ to go above $m$.

In Section 5.1, we consider an auxiliary boundary value problem, introduce the critical high-water mark $m^*$ and prove item (i) above. In Section 5.2,  we consider a related optimal controller-stopper problem, show that its solution is the Legendre transform of the minimum probability of drawdown when $m\le m^*$, and prove item (ii) above. Finally, Section 5.3 provides further properties of the optimal investment strategy.

\subsect{5.1  Minimum probability of drawdown when $m^* < m < c/r$}

When wealth reaches the initial maximum wealth $M_0 = m$, the individual either allows wealth to increase above this level or does not. In this section, we identify values of $m \in (0,c/r)$ for which it is optimal to increase maximum wealth.

%


For an arbitrary constant $m_0 \in (0, c/r)$, consider the following boundary-value problem (BVP). For $m_0 \le m \le c/r$ and $\al m \le w \le m$,
\begin{equation}\label{eq:BVP1}
	\left\{
	\eqalign{
	& \la h = (r w - c) h_w + \min_\pi \left[ (\mu - r) \pi h_w + {1 \over 2} \sigma^2 \pi^2 h_{ww} \right], \cr
	&h(\al m, m) = 1,  \quad h_m(m, m) = 0, \cr
	&\lim_{m \to c/r-} h \left(m, m \right)= 0.}
	\right.
\end{equation}
According to Corollary 3.2, if we find a classical solution of BVP \eqref{eq:BVP1} that is non-increasing and convex with respect to $w$, then that solution equals the minimum probability of drawdown $\phi$ for $m_0 \le m \le c/r$ and $\al m \le w \le m$. Note that, we include $m = c/r$ in the boundary-value problem \eqref{eq:BVP1} so that we can use information about $\phi$ at this point.


To find the desired solution of BVP \eqref{eq:BVP1}, we first solve a related free-boundary problem; then, we show that its convex dual via the Legendre transform solves \eqref{eq:BVP1} and, thus, equals the minimum probability of lifetime drawdown. Consider the following free-boundary problem (FBP) on $(y,m)\in[\tilde{y}_m(m), \tilde{y}_{\al m}(m)]\times[m_0,c/r]$, with $0 < \tilde{y}_m(m) < \tilde{y}_{\al m}(m)$ to be determined.
\begin{equation}\label{eq:FBP1}
\left\{
\eqalign{
&\del y^2 \tilde \phi_{yy} - (r - \la) y \tilde \phi_y - \la \tilde \phi + c y = 0, \cr
&\tilde \phi(\tilde{y}_{\al m}(m), m) = 1 + \al m \, \tilde{y}_{\al m}(m), \quad \tilde \phi_y(\tilde{y}_{\al m}(m), m) = \al m, \cr
&\tilde \phi_y(\tilde{y}_m(m), m) = m, \qquad \tilde \phi_{m}(\tilde{y}_m(m), m) = 0, \cr
&\lim_{m \to c/r-} \tilde \phi \left(\tilde{y}_m(m), m \right) = {c \over r} \, \tilde{y}_m\left({c \over r} \right), \qquad \lim_{m \to c/r-} \tilde \phi_y\left(\tilde{y}_m(m), m \right) = {c \over r}.}
\right.
\end{equation}
In the following proposition, we present the solution of the FBP \eqref{eq:FBP1}.

\prop{5.1}{ For a given constant $m_0\in(0,c/r)$ and functions $g_0$, $g_1$, $h_0$, $h_1$, and $h_2$ explicitly given in Appendix A, consider the following non-linear first order ODE
\begin{equation}\label{eq:ODE1}
z^\prime(m) = {g_1(z)(c/r-m) + g_0(z) \over h_2(z)(c/r-m)^2+h_1(z)(c/r-m)+h_0(z)};\quad m_0 \le m < c/r.
\end{equation}
Assume that \eqref{eq:ODE1} has a classical solution $z: [m_0,c/r]\to [0, 1]$ satisfying the terminal condition $z(c/r) = 0$. Then, a solution of FBP \eqref{eq:FBP1} for $(y,m)\in[\tilde{y}_m(m), \tilde{y}_{\al m}(m)]\times[m_0,c/r]$ is given by
\begin{equation}\label{eq:phiTilde}
\eqalign{
\tilde \phi(y, m) = {c \over r} \, y &- \left[ {B_2 \over B_1 - B_2} + \left( {c \over r} - \al m \right) {1 - B_2 \over B_1 - B_2} \, \tilde{y}_{\al m}(m) \right] \left( {y \over \tilde{y}_{\al m}(m)} \right)^{B_1}\cr
&+ \left[ {B_1 \over B_1 - B_2} - \left( {c \over r} - \al m \right) {B_1 - 1 \over B_1 - B_2} \, \tilde{y}_{\al m}(m) \right] \left( {y \over \tilde{y}_{\al m}(m)} \right)^{B_2},
}
\end{equation}
in which
\begin{align}
	B_1 &= {1 \over 2 \del} \left[ (r - \la + \del) + \sqrt{(r - \la + \del)^2 + 4 \la \del} \right] = {\gam \over \gam - 1} > 1, \label{eq:B1}\\
	B_2 &= {1 \over 2 \del} \left[ (r - \la + \del) - \sqrt{(r - \la + \del)^2 + 4 \la \del} \right] < 0, \label{eq:B2}
\end{align}
the free boundary $\tilde{y}_{\al m}(m)$ is given in terms of $z(m)$ by
\begin{equation}\label{eq:yalm1}
\eqalign{
& {1 \over \tilde{y}_{\al m}(m)} \, {B_1 B_2 \over B_1 - B_2} \left( z(m)^{B_1-1} - z(m)^{B_2-1} \right)  \cr
& \quad = \left( {c \over r} - m \right) - \left( {c \over r} - \al m \right) \left[ {B_1 (1 - B_2) \over B_1 - B_2} z(m)^{B_1-1} + {B_2(B_1 - 1) \over B_1 - B_2} z(m)^{B_2-1} \right],}
\end{equation}
and the free boundary $\tilde{y}_m(m)\in\big(0,\tilde{y}_{\al m}(m)\big)$ is given by $\tilde{y}_m(m) = \tilde{y}_{\al m}(m) \, z(m)$. Furthermore, for all $m\in[m_0,c/r]$, $\tilde\phi (\cdot,m)$ is increasing and concave on $[\tilde{y}_m(m), \tilde{y}_{\al m}(m)]$. \qed}

\pf The general solution to FBP \eqref{eq:FBP1} is given by
\begin{equation}\label{eq:Ansatz}
\tilde \phi(y, m) = \tilde D_1(m) \, y^{B_1} + \tilde D_2(m) \, y^{B_2} + {c \over r} \, y,
\end{equation}
in which $B_1$ and $B_2$ are given in \eqref{eq:B1} and \eqref{eq:B2}, respectively, and $\tilde D_1(m)$ and $\tilde D_2(m)$ are functions of $m$ to be determined.  The boundary conditions of \eqref{eq:FBP1} imply that
\begin{equation}\label{eq:BC1}
\tilde D_1(m) \, \tilde{y}_{\al m}(m)^{B_1} + \tilde D_2(m) \, \tilde{y}_{\al m}(m)^{B_2} + {c \over r} \, \tilde{y}_{\al m}(m) = 1 + \al m \, \tilde{y}_{\al m}(m),
\end{equation}
\begin{equation}\label{eq:BC2}
\tilde D_1(m) \, B_1 \, \tilde{y}_{\al m}(m)^{B_1 - 1} + \tilde D_2(m) \, B_2 \, \tilde{y}_{\al m}(m)^{B_2 - 1} + {c \over r} = \al m,
\end{equation}
\begin{equation}\label{eq:BC3}
(\tilde D_1)'(m) \, \tilde{y}_m(m)^{B_1} + (\tilde D_2)'(m) \, \tilde{y}_m(m)^{B_2} = 0,
\end{equation}
\begin{equation}\label{eq:BC4}
\tilde D_1(m) \, B_1 \, \tilde{y}_m(m)^{B_1 - 1} + \tilde D_2(m) \, B_2 \, \tilde{y}_m(m)^{B_2 - 1} + {c \over r} = m,
\end{equation}
\begin{equation}\label{eq:BC5}
\lim_{m \to c/r-} \left[\tilde D_1(m) \, \tilde{y}_m(m)^{B_1} + \tilde D_2(m) \, \tilde{y}_m(m)^{B_2} \right] = 0,
\end{equation}
and
\begin{equation}\label{eq:BC6}
\lim_{m \to c/r-} \left[ \tilde D_1(m) \, B_1 \, \tilde{y}_m(m)^{B_1-1} + \tilde D_2(m) \, B_2 \, \tilde{y}_m(m)^{B_2-1} \right] = 0.
\end{equation}

Solve equations \eqref{eq:BC1} and \eqref{eq:BC2} for $\tilde D_1(m)$ and $\tilde D_2(m)$ to get
\begin{equation}\label{eq:D1}
\tilde D_1(m) = - {B_2 \over B_1 - B_2} \, \tilde{y}_{\al m}(m)^{-B_1} -  {1 - B_2 \over B_1 - B_2} \left( {c \over r} - \al m \right) \tilde{y}_{\al m}(m)^{1- B_1},
\end{equation}
and
\begin{equation}\label{eq:D2}
\tilde D_2(m) =  {B_1 \over B_1 - B_2} \, \tilde{y}_{\al m}(m)^{-B_2} -  {B_1 - 1 \over B_1 - B_2} \left( {c \over r} - \al m \right) \tilde{y}_{\al m}(m)^{1 - B_2}.
\end{equation}
Substituting these expressions into equations \eqref{eq:Ansatz} yields \eqref{eq:phiTilde}. Similarly, substituting into \eqref{eq:BC4} yields \eqref{eq:yalm1}, in which we define $z(m) :=  \tilde{y}_m(m)/\tilde{y}_{\al m}(m)$.

Next, differentiate \eqref{eq:D1} and \eqref{eq:D2} with respect to $m$ and substitute the results into equation \eqref{eq:BC3} to get
$$
\eqalign{
& {\tilde{y}'_{\al m}(m) \over \tilde{y}_{\al m}(m)} \left( z(m)^{B_1-1} - z(m)^{B_2-1} \right) \left[ {B_1 B_2 \over \tilde{y}_{\al m}(m)} + (B_1 - 1)(1 - B_2)  \left( {c \over r} - \al m \right) \right] \cr
& \quad + \al \left( (1 - B_2) \, z(m)^{B_1-1} + (B_1 - 1) \, z(m)^{B_2-1} \right) = 0. }
$$
By replacing $\tilde{y}_{\al m}$ from \eqref{eq:yalm1} while keeping $\tilde{y}'_{\al m}/\tilde{y}_{\al m}$, we obtain (after some rearranging)
\begin{equation}\label{eq:Aux1}
{\tilde{y}'_{\al m} \over \tilde{y}_{\al m}} = {- \al \left[ (1-B_2) \, z(m)^{B_1-1} + (B_1-1) \, z(m)^{B_2-1} \right] \over (B_1-B_2) \left( {c \over r} - m \right) - \left( {c \over r} - \al m \right) \left[ (1-B_2) \, z(m)^{B_1-1} + (B_1-1) \, z(m)^{B_2-1} \right] }.
\end{equation}
By differentiating \eqref{eq:yalm1} with respect to $m$ and eliminating $\tilde{y}'_{\al m}/ \tilde{y}_{\al m}$ from the resulting equation via \eqref{eq:Aux1}, it follows that $z(m)$ satisfies ODE \eqref{eq:ODE1} for $m\in[m_0,c/r)$.

We now reverse the above argument and conclude that if $z(m)$ satisfies ODE \eqref{eq:ODE1} with the terminal condition $z(c/r) = 0$, with $\tilde{y}_{\al m}$ given by \eqref{eq:yalm1}, and with $\tilde{y}_m(m) = z(m) \, \tilde{y}_{\al m}(m)$, then $\tilde \phi(y, m)$ given by \eqref{eq:phiTilde} satisfies FBP \eqref{eq:FBP1}.  Monotonicity and concavity of $\tilde \phi$ with respect to $y$ can be easily shown by differentiating \eqref{eq:phiTilde}.

It only remains to check the boundary conditions at $m = c/r$, that is, \eqref{eq:BC5} and \eqref{eq:BC6}. \eqref{eq:BC6} directly follows from \eqref{eq:BC4}. Furthermore, \eqref{eq:BC5} holds if $\lim_{m\to c/r^-} z(m) = 0$. To show this, note that by \eqref{eq:BC4} and \eqref{eq:D1}
\begin{equation}\label{eq:Aux2}
\eqalign{
	\tilde D_1(m) \, \tilde{y}_m(m)^{B_1} &+ \tilde D_2(m) \, \tilde{y}_m(m)^{B_2}\cr
	&= {1 \over B_2}(m-c/r) \, \tilde{y}_m(m) + \left(1-{B_1\over B_2} \right) \tilde D_1(m) \, \tilde{y}_m(m)^{B_1}\cr
	&= {1\over B_2} (m-c/r)\, z(m) \, \tilde{y}_{\al m}(m) \cr
	&\qquad + \left[1+{1-B_2\over B_2}(c/r-\alpha m) \tilde{y}_{\al m}(m) \right] z(m)^{B_1} .
}
\end{equation}
By imposing the boundary condition $\lim_{m\to c/r^-} z(m) = 0$ in \eqref{eq:yalm1}, we obtain $\lim_{m\to c/r^-}\tilde{y}_{\al m}(m) = {B_1 \over (B_1-1)(1-\al)c/r}$. Thus, taking the limit of \eqref{eq:Aux2} as $m\to c/r^-$ yields \eqref{eq:BC5}.  \qed

The main assumption of Proposition 5.1 is the existence of a solution to ODE \eqref{eq:ODE1} satisfying the terminal condition $z(c/r)=0$. One can show that the right side of ODE \eqref{eq:ODE1} is not continuous at $(c/r,0)$; thus, existence of such a solution is not trivial. Proposition 5.4 below provides conditions under which such a solution exists.

Furthermore, we have not yet specified the value of $m_0$. According to Proposition 5.1, a solution $z$ which exists on an interval $[m_0,c/r]$ yields a solution of FBP \eqref{eq:FBP1} on the same interval.  Naturally, we are interested in the smallest $m_0 \in (0,c/r)$ for which such solution exists.  We denote this smallest value by $m^*$. Proposition 5.4 also identifies $m^*$.

Before providing the main result on the existence of a solution to ODE \eqref{eq:ODE1}, we introduce some preliminary results and definitions.  First, we introduce a function $x(m)$ and a constant $\widehat{m} \in (0,c/r)$.  

\lem{5.2}{There exists an increasing function $x:[0,c/r)\to[1,+\infty)$ which uniquely solves
\begin{equation}\label{eq:xm}
\left( {c \over r} - m \right) \left[ {1 - B_2 \over B_1 - B_2} \, x(m)^{B_1 - 1} + {B_1 - 1 \over B_1 - B_2} \, x(m)^{B_2 - 1} \right] = {c \over r} - \al m.
\end{equation}
}

\pf The left side of \eqref{eq:xm} increases with respect to $x(m) > 1$; as $x(m)$ approaches $1+$, the left side of \eqref{eq:xm} approaches $c/r - m$, which is less than $c/r - \al m$; and, as $x(m)$ approaches $\infty$, the left side approaches $\infty$. Thus, \eqref{eq:xm} has a unique solution. Differentiating \eqref{eq:xm} with respect to $m$ implies that $x(m)$ is increasing with $m$.
\qed

\lem{5.3}{There exists a unique solution $\widehat{m}\in \left({c \over r} \left( 1 + {1 - \al \over \al B_2} \right)_+, \; {c \over r} \right)$ of the following equation:
\begin{equation}\label{eq:mhat}
\left[ \al B_1 + {{c \over r}(1 - \al) \over {c \over r} - \widehat{m}} \right]^{{1 \over B_1 - 1}} = \left[ \al B_2 + {{c \over r}(1 - \al) \over {c \over r} - \widehat{m}} \right]^{-{1 \over 1 - B_2}}.
\end{equation}
}

\pf Define $g$ by
$$
g(m) = \left[ \al B_1 + {{c \over r}(1 - \al) \over {c \over r} - m} \right]^{{1 \over B_1 - 1}} - \left[ \al B_2 + {{c \over r}(1 - \al) \over {c \over r} - m} \right]^{-{1 \over 1 - B_2}},
$$
for $m \in \left({c \over r} \left( 1 + {1 - \al \over \al B_2} \right)_+, \; {c \over r} \right)$, and note that $g$ is increasing with $m$.

We have two cases to consider.  First, if $1 - \al(1 - B_2) \le 0$, then $g$ increases from $-\infty$ to $\infty$ as $m$ increases from ${c \over r} \left( 1 + {1 - \al \over \al B_2} \right) \ge 0$ to ${c \over r}$.  Thus, $g$ has a unique zero, $\widehat{m}> 0$, in this interval.

Second, if $1 - \al(1 - B_2) > 0$, then to show that $g$ has a unique zero in $(0, c/r)$, it is enough to show that $g(0) < 0$.  To this end, note that $g(0) < 0$ is equivalent to
$$
(1 + x)^{{1 \over x}} < (1 - y)^{-{1 \over y}},
$$
in which $x = \al(B_1 - 1) > 0$ and $y = \al(1 - B_2) > 0$.  This inequality holds because the left side is less than $e$ and the right side is greater than $e$.   \qed

\begin{figure}[t]
\centerline{
\adjustbox{trim={0.05\width} {0.0\height} {0.04\width} {0.06\height},clip}{\includegraphics[scale=0.35,page=1]{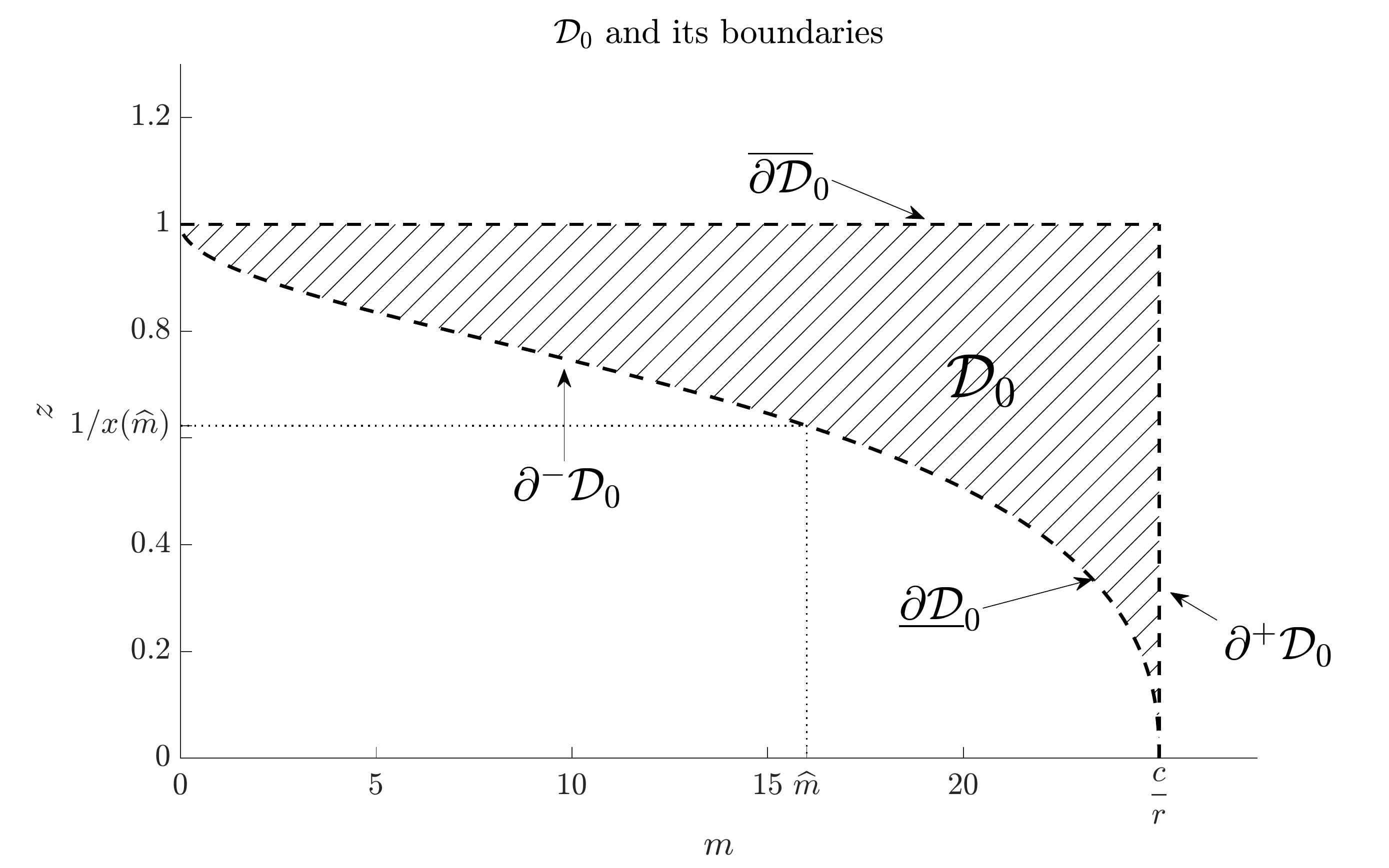}}
\hspace{0.25em}
\adjustbox{trim={0.05\width} {0.0\height} {0.04\width} {0.06\height},clip}{\includegraphics[scale=0.35,page=1]{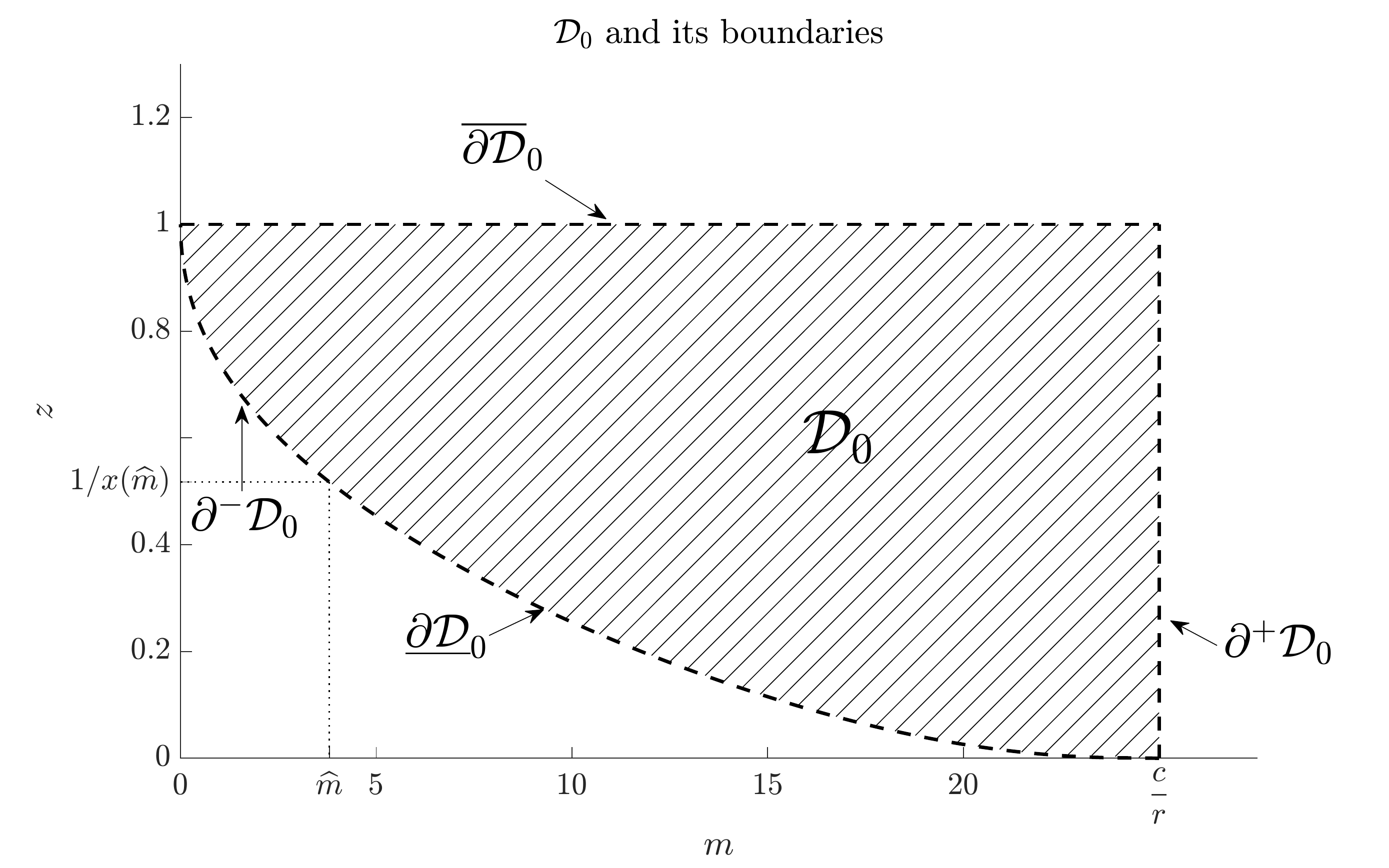}}
}
\nobreak
\centerline{{\bf Figure 1}: The Domain $\mathcal{D}_0$ and its boundaries for two sets of parameters. On the left, $\mu = 0.06$, $\sigma = 0.20$, }
\centerline{$r = 0.04$, $c = 1$, $\lambda = 0.04$, and $\alpha = 0.50$. On the right, $\mu=0.12$, with the other parameters unchanged.}
\end{figure}

Given $x(m)$ and $\widehat{m}$, we define the region $\mathcal{D}_0\subset[0,c/r]\times[0,1]$ by
\begin{equation}\label{eq:D0}
	\mathcal{D}_0:=\{(m,z):0\le m\le c/r,\frac{1}{x(m)}\le z \le 1\},
\end{equation}
and also its upper, lower, right, and left boundaries, by
\[
\begin{split}
	\overline{\partial\mathcal{D}}_0&=\Big\{(m,1):0\le m\le c/r\Big\},\\
	\underline{\partial\mathcal{D}}_0&=\Big\{\big(m,1/x(m)\big):\widehat{m}< m\le c/r\Big\},\\
	\partial^+\mathcal{D}_0&= \Big\{(c/r,z):0< z< 1\Big\},
\end{split}
\]
and
\[
	\partial^-\mathcal{D}_0=\Big\{\big(m,1/x(m)\big):0\le m\le \widehat{m}\Big\},
\]
respectively. Figure 1 illustrates $\mathcal{D}_0$ and its boundaries for two sets of parameters. Specifically, for the left graph, we chose $\mu = 0.06$, $\sigma = 0.20$, $r = 0.04$, $c = 1$, $\lambda = 0.04$, and $\alpha = 0.50$; and for the graph on right, we chose $\mu = 0.12$, with the other parameters unchanged.  We will use the same sets of parameters for later illustrations.

See Appendix B for a proof that the right side of ODE \eqref{eq:ODE1} is continuous in the interior of $\mathcal{D}_0$, as well as on the upper-right boundary $\overline{\partial\mathcal{D}}_0\cup \partial^+\mathcal{D}_0$. Furthermore, the right side of ODE \eqref{eq:ODE1} approaches $\pm \infty$ on the lower-left boundary $\underline{\partial\mathcal{D}}_0\cup \partial^-\mathcal{D}_0$, except at singularities $\big(\widehat{m},1/x(\widehat{m})\big)$ and $(0,c/r)$, where it has no limit. Thus, by classical existence uniqueness theorems, there exists a unique solution passing through any point in $\mathcal{D}_0\backslash\{\big(\widehat{m},1/x(\widehat{m})\big),(c/r,0)\}$; see Lemma B.4.

\begin{figure}[t]
\centerline{
\adjustbox{trim={0.015\width} {0.0\height} {0.12\width} {0.06\height},clip}{\includegraphics[scale=0.33,page=1]{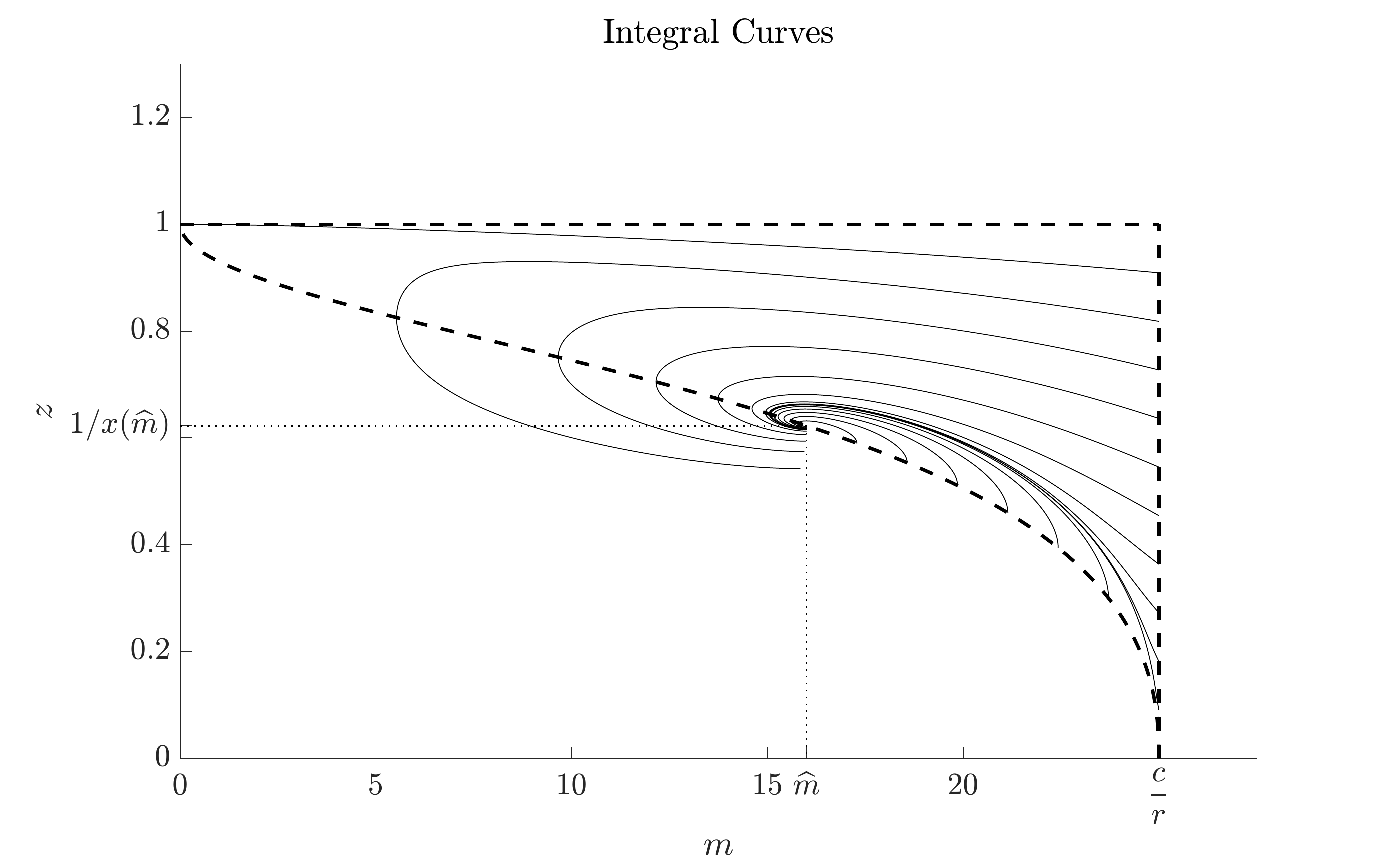}}
\hspace{1em}
\adjustbox{trim={0.015\width} {0.0\height} {0.05\width} {0.06\height},clip}{\includegraphics[scale=0.33,page=1]{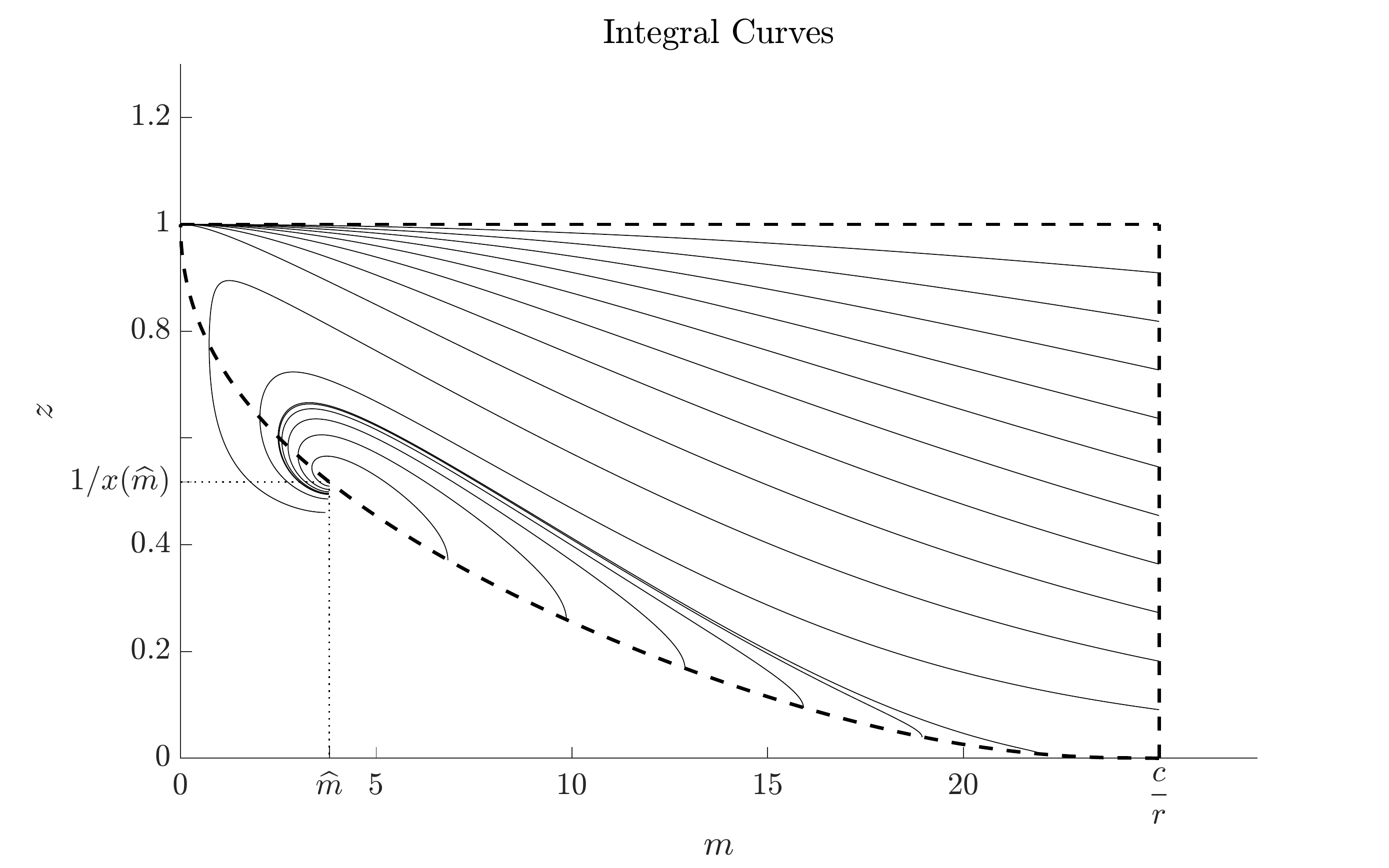}}
}
\nobreak
\centerline{{\bf Figure 2}: Integral curves of ODE \eqref{eq:ODE1} passing through points in $\underline{\partial\mathcal{D}}_0\cup\partial^+\mathcal{D}_0\backslash\{(c/r,0)\}$.}
\end{figure}

Figure 2 illustrates integral curves of ODE \eqref{eq:ODE1} passing through points in $\underline{\partial\mathcal{D}}_0\cup\partial^+\mathcal{D}_0\backslash\{(c/r,0)\}$. We used the MATLAB ODE solver ``ode45'' to numerically approximate the integral curves. As an aside, the right side of ODE \eqref{eq:ODE1} becomes $-\infty$ at the lower boundary $\underline{\partial\mathcal{D}}_0\backslash\{(c/r,0)\}$. In such cases, it is more robust to solve the Abel equation
\begin{equation}\label{eq:Abel}
\left\{
\begin{split}
	&m^\prime(z) = {h_2(z)(c/r-m)^2+h_1(z)(c/r-m)+h_0(z) \over g_1(z)(c/r-m) + g_0(z)}, \quad \hbox{ for } (z,m(z))\in \mathcal{D}_0^\top,\\
	&m(z_0)=m_0,
\end{split}
\right.
\end{equation}
and invert the solution to get the solution of \eqref{eq:ODE1}.  Then, solve \eqref{eq:ODE1} directly near the points where the right side of \eqref{eq:Abel} become unbounded. Finally, create the integral curves by pasting together the solutions thus found. 

The integral curves passing through points near $(c/r,0)$ spiral around the singularity $\big(\widehat{m},1/x(\widehat{m})\big)$. In particular, a curve passing through a given point $(m_0, z_0)$ in $\underline{\partial\mathcal{D}}_0\cup\partial^+\mathcal{D}_0\backslash\{(c/r,0)\}$ near $(c/r,0)$ hits the graph of $z = 1/x(m)$ on a point $\big(\widetilde{m},1/x(\widetilde{m})\big)$, for some $0 < \widetilde{m} < \widehat{m}$. Then, the integral curve spirals back towards the singularity $\big(\widehat{m},1/x(\widehat{m})\big)$. Thus, such an integral curve can only be defined on the interval $[\widetilde{m}, m_0]$. In particular, it is not defined on the interval $(0,\widetilde{m})$.

The solution of ODE \eqref{eq:ODE1} satisfying the terminal condition $z(c/r) = 0$ inherits this behavior.  We denote by $m^*<\widehat{m}$, the $m$-value where the solution intercepts the left boundary $\partial^-\mathcal{D}_0$.  That is, $m^*$ is the value of $\widetilde{m}$ from the above paragraph that corresponds to $(m_0, z_0) = (c/r^-, 0^+)$.  We demonstrate these assertions in the next proposition.

\prop{5.4}{Assume that there exist solutions $\underline{z}(m)$ and $\overline{z}(m)$ of \eqref{eq:ODE1} in $\mathcal{D}_0$ such that $\underline{z}$ (respectively, $\overline{z}$) satisfies the terminal condition $z(m_0)=z_0$ for $(m_0,z_0)\in\underline{\partial\mathcal{D}}_0$ (respectively, $(m_0,z_0)\in\partial^+\mathcal{D}_0$) and extends on the left to $\partial^-\mathcal{D}_0\backslash\big\{\big(\widehat{m},1/x(\widehat{m})\big)\big\}$. Let $\overline{m}^*$ (respectively, $\underline{m}^*$) be the value of $m$ where $\underline{z}$ (respectively, $\overline{z}$) intercepts $\partial^-\mathcal{D}_0$. Then, there exists a unique solution $z(m)$ of \eqref{eq:ODE1} in $\mathcal{D}_0$ satisfying the terminal condition $z(c/r) = 0$ and extending on the left to the boundary $\partial^-\mathcal{D}_0$ such that $z(m^*)=1/x(m^*)$ for some $m^*\in(\underline{m}^*,\overline{m}^*)$. In particular, $z(m)$ is not defined on $(0, m^*)$.
}

\pf See Appendix B.\qed

Proposition 5.4 implies that there is a unique solution of \eqref{eq:ODE1} satisfying $z(c/r) = 0$.  Figure 2 provides several candidates for $\underline{z}(m)$ and $\overline{z}(m)$ in our numerical example.  We can closely approximate these solutions by solving \eqref{eq:ODE1} with the terminal condition $z(c/r)=\epsilon$, for a small number $\epsilon>0$. Such approximations are illustrated in Figure 3.

\begin{figure}[ht]
\centerline{
\adjustbox{trim={0.015\width} {0.0\height} {0.12\width} {0.06\height},clip}{\includegraphics[scale=0.35,page=1]{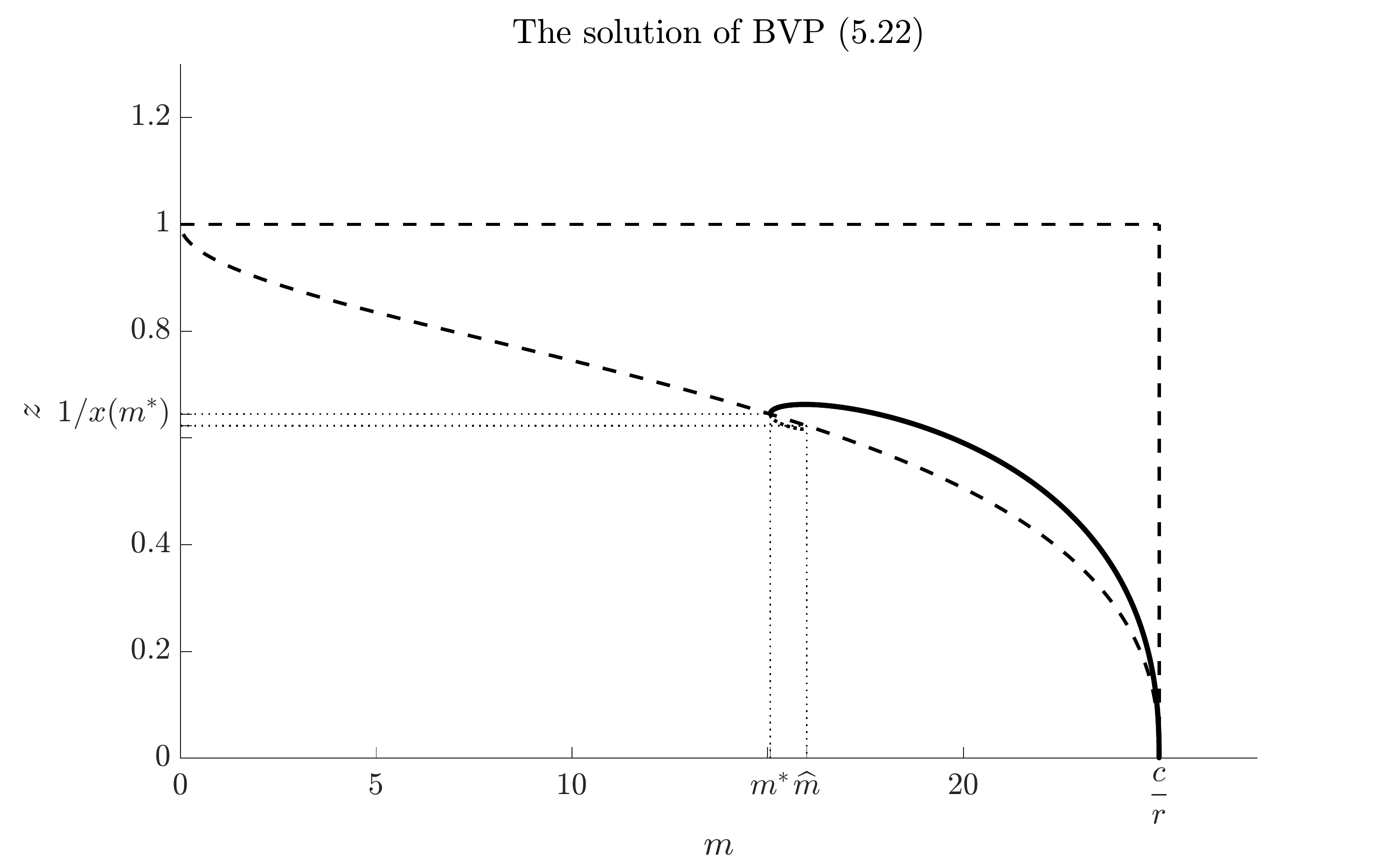}}
\hspace{1em}
\adjustbox{trim={0.015\width} {0.0\height} {0.05\width} {0.06\height},clip}{\includegraphics[scale=0.35,page=1]{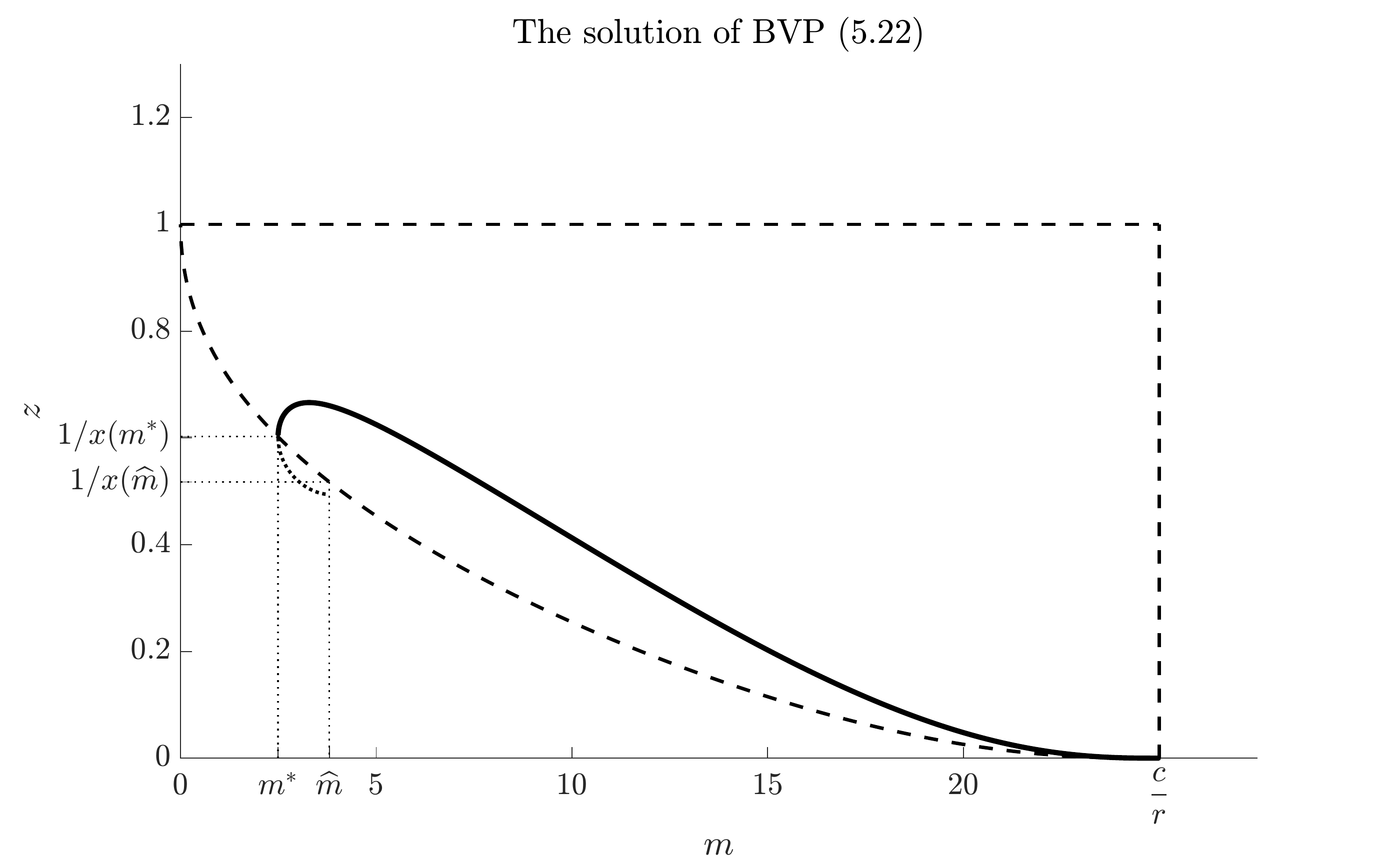}}
}
\nobreak
\centerline{{\bf Figure 3}: The solution of ODE \eqref{eq:ODE1} satisfying $z(c/r)=0$.}\vspace{1em}
\end{figure}

Once a solution of ODE \eqref{eq:ODE1} satisfying $z(c/r) = 0$ is found, it is straightforward to find the minimum probability of drawdown for $m^*\le m\le c/r$. Indeed, by Proposition 5.1, such solution $z$ yields a solution $\tilde \phi$ of FBP \eqref{eq:FBP1} for $m^*\le m\le c/r$. Because $\tilde \phi$ in \eqref{eq:ODE1} is concave with respect to $y$, we can define its convex dual $\Psi$ via the Legendre transform. The following result states that $\Psi$ is the minimum probability of drawdown for $m^*\le m\le c/r$. 

\prop{5.5} {Define $\Psi$ by
\begin{equation}\label{eq:Psi}
\Psi(w, m) = \max_y \left( \tilde \phi(y, m) - w y \right),
\end{equation}
in which $\al m \le w \le m$ and $m^* \le m \le c/r$.  Then, the minimum probability of lifetime drawdown equals $\Psi$ when $m^* \le m \le c/r$.}

\pf From \eqref{eq:Psi}, it follows that the critical value $y^*$ solves $w = \tilde \phi_y(y, m)$; thus, given $w$, we have $y^* = I(w, m)$, in which $I$ is the  functional inverse of $\tilde \phi_y$ with respect to $y$.  Therefore, $\Psi(w, m) = \tilde \phi(I(w, m), m) - w I(w, m)$.  By differentiating this expression of $\Psi$ with respect to $w$, we obtain $\Psi_w(w, m) = \tilde \phi_y(I(w, m), m) \, I_w(w, m) - I(w, m) - w \, I_w(w, m) = - I(w, m)$; thus, $y^* = - \Psi_w(w, m)$.  Also, note that $\Psi_{ww}(w, m) = -1/\tilde \phi_{yy}(I(w, m), m)$.

By substituting $y^* = - \Psi_w(w, m)$ into the free-boundary problem for $\tilde \phi$, namely \eqref{eq:FBP1}, we learn that $\Psi$ solves the boundary-value problem \eqref{eq:BVP1}. Moreover, because $\tilde \phi$ is increasing and concave with respect to $y$, $\Psi$ is decreasing and convex with respect to $w$.  Thus, Corollary 3.2 implies that $\Psi$ equals the minimum probability of lifetime drawdown when $m^* \le m \le c/r$. \qed

We have a theorem that follows immediately from Proposition 5.5.

\th{5.6} {Assume that the conditions of Proposition 5.4 hold such that $z(m)$ is the solution of ODE \eqref{eq:ODE1} on $[m^*,c/r]$ satisfying $z(c/r)=0$. Furthermore, define $\tilde{y}_{\al m}(m)$ by \eqref{eq:yalm1} and $\tilde{y}_m(m)=z(m) \, \tilde{y}_{\al m}(m)$. Then, for $\al m \le w \le m$ and $m^* \le m \le c/r$, the minimum probability of lifetime drawdown $\phi$ is given by
\begin{equation}\label{eq:phi1}
\eqalign{
\phi(w, m) &= {B_1 - 1 \over B_1 - B_2} \left[ B_2 + \left( {c \over r} - \al m \right) (1 - B_2) \, \tilde{y}_{\al m}(m) \right] \left( {y \over \tilde{y}_{\al m}(m)} \right)^{B_1} \cr
& \quad + {1 - B_2 \over B_1 - B_2} \left[ B_1 - \left( {c \over r} - \al m \right) (B_1 - 1) \, \tilde{y}_{\al m}(m) \right] \left( {y \over \tilde{y}_{\al m}(m)} \right)^{B_2},
}
\end{equation}
in which $y \in [\tilde{y}_m(m), \tilde{y}_{\al m}(m)]$ uniquely solves
\begin{equation}\label{eq:I1}
\eqalign{
{c \over r} - w &=  {B_1  \over B_1 - B_2} \left[ { B_2 \over  \tilde{y}_{\al m}(m)} + \left( {c \over r} - \al m \right) (1 - B_2) \right] \left( {y \over \tilde{y}_{\al m}(m)} \right)^{B_1 - 1} \cr
& \quad - {B_2 \over B_1 - B_2} \left[ {B_1 \over  \tilde{y}_{\al m}(m)} - \left( {c \over r} - \al m \right) (B_1 - 1) \right] \left( {y \over \tilde{y}_{\al m}(m)} \right)^{B_2 - 1}.
}
\end{equation}
For wealth between $\al m$ and $m$, the corresponding optimal investment strategy $\pi^*$ is given in feedback form by $\pi^*_t = \pi^*(W^*_t, M^*_t)$, in which
\begin{equation}\label{eq:piStar1}
\eqalign{
\pi^*(w, m) &= {\mu - r \over \sigma^2} \,  {B_1 (B_1 - 1) \over B_1 - B_2} \left[  {B_2 \over \tilde{y}_{\al m}(m)} +  \left( {c \over r} - \al m \right) (1 - B_2)  \right] \left( {y \over \tilde{y}_{\al m}(m)} \right)^{B_1 - 1} \cr
& \quad +  {\mu - r \over \sigma^2} \,  {B_2 (1 - B_2) \over B_1 - B_2} \left[  {B_1 \over \tilde{y}_{\al m}(m)} -  \left( {c \over r} - \al m \right) (B_1 - 1)  \right]  \left( {y \over \tilde{y}_{\al m}(m)} \right)^{B_2 - 1}.}
\end{equation}
 \qed}
 
Figure 4 illustrates the optimal investment in the risky asset when $m^* \le w = m \le c/r$, that is, when the wealth reaches the high-water mark, obtained from \eqref{eq:piStar1}. As expected from Section 4, $\pi^*(c/r,c/r) = 0$, that is, the optimal allocation when wealth reaches the safe level $c/r$ is not to invest in the stock. For $m^*< m < c/r$, we have $\pi^*(m,m) > 0$, which means that for these values of $m$ the optimal allocation allows the high-water mark to increase. Finally, at $w=m=m^*$, we have $\pi^*(m^*,m^*) = 0$. Since the consumption rate $c$ is larger than the riskless return $r\,m^*$, wealth can never become larger than $m^*$. In other words, the optimal allocation for $m = m^*$ does not let the high-water mark  increase.  In the next section, we will show that for all values $0 < m < m^*$, it is optimal not to let the maximum wealth to increase.
 
 \begin{figure}[t]
 \centerline{
 \adjustbox{trim={0.55\width} {0.7\height} {0.095\width} {0.05\height},clip}{\includegraphics[scale=0.45,page=1]{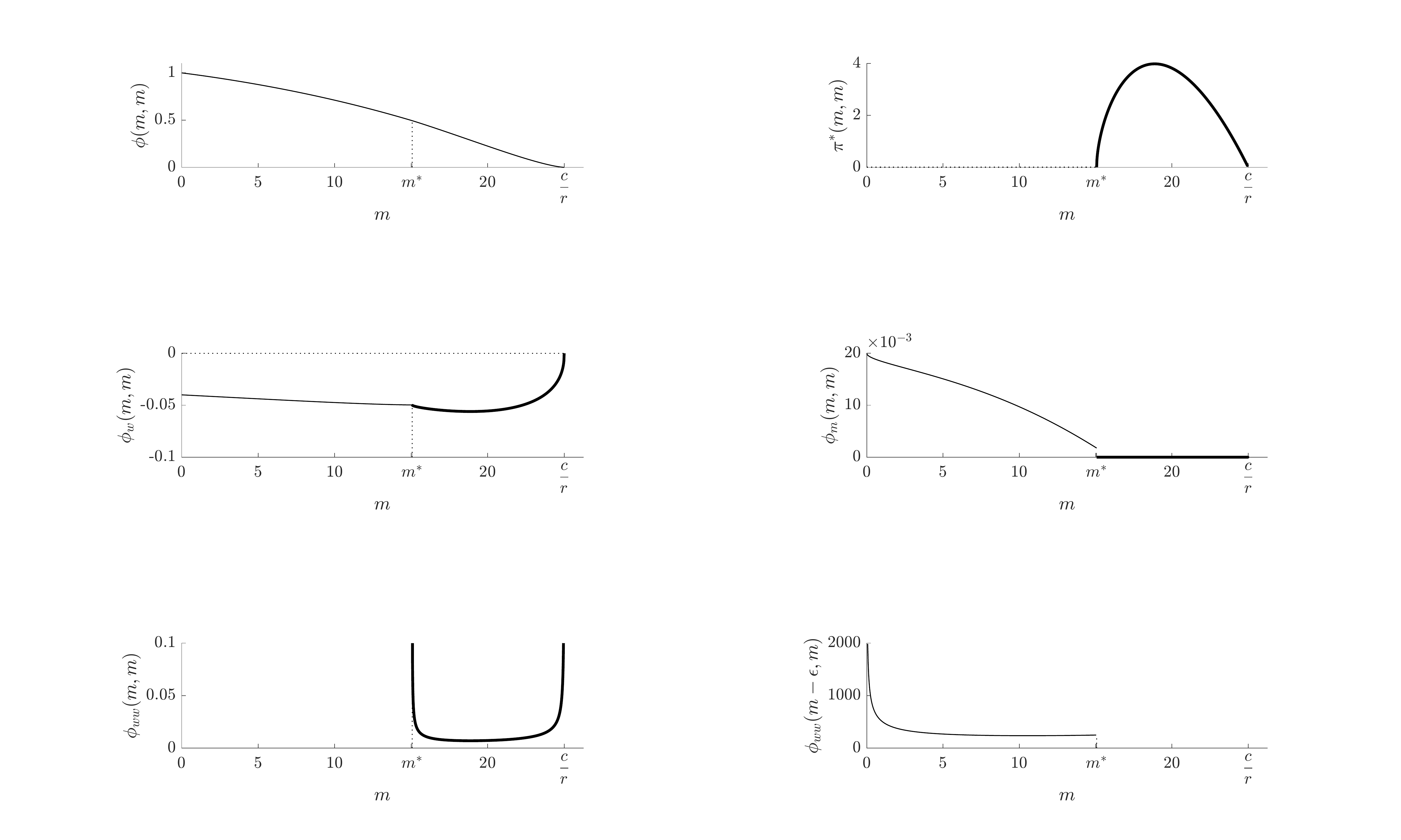}}
 \hspace{2em}
 \adjustbox{trim={0.55\width} {0.7\height} {0.095\width} {0.05\height},clip}{\includegraphics[scale=0.45,page=1]{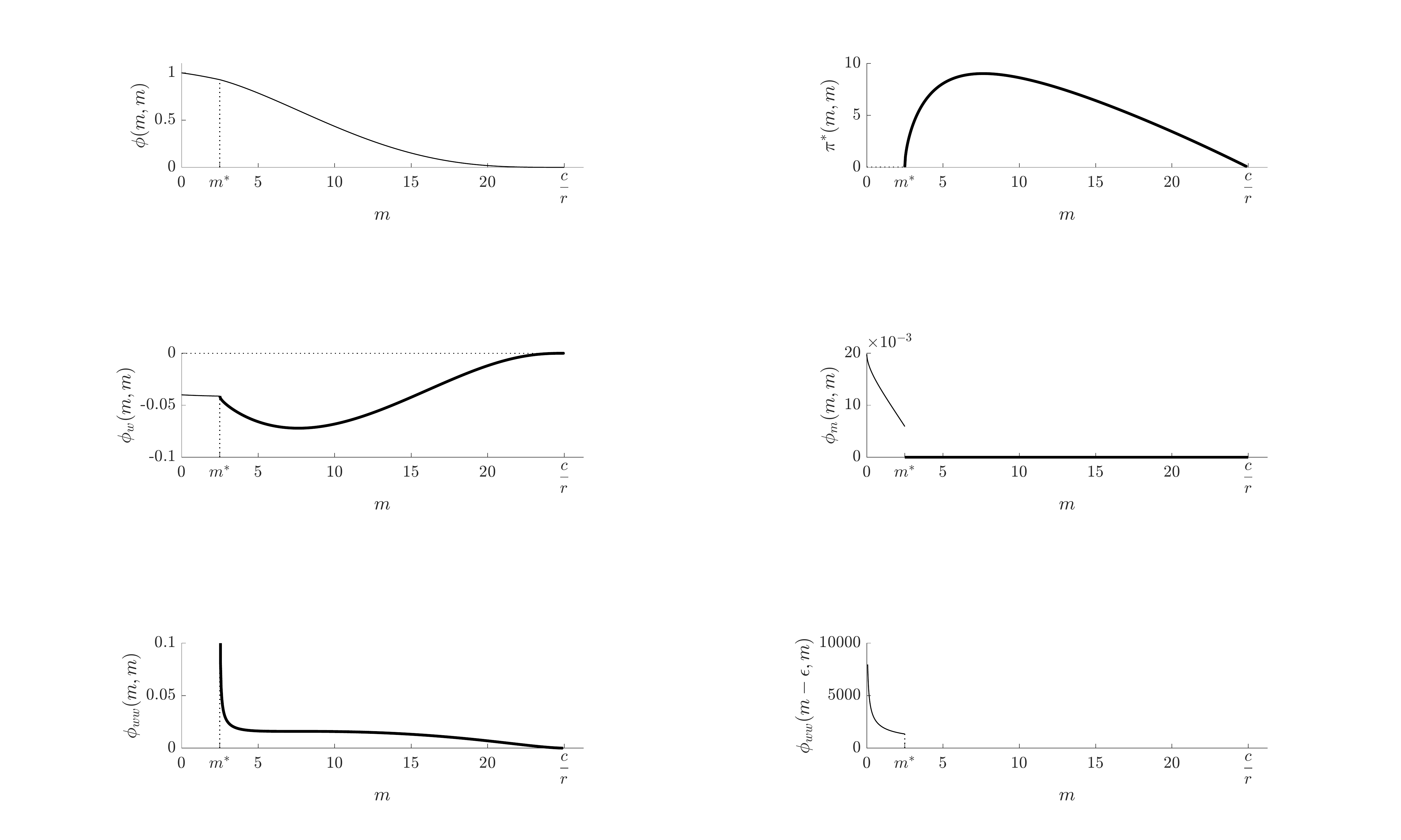}}
 }
 \nobreak
 \centerline{{\bf Figure 4}: Optimal investment in the risky asset when the high-water mark is reached,}
 \centerline{that is, $\pi^*(m,m) =  -{\mu - r \over \sigma^2} \, {\phi_w(m,m) \over \phi_{ww}(m,m)}$, for $m^*<m<c/r$.}\vspace{1em}
 \end{figure}

\subsect{5.2 Minimum probability of drawdown when $0 < m < m^*$}

In this section, we first study a related optimal controller-stopper problem and, then, show that its solution is the Legendre transform of the minimum probability of drawdown when $0<m\le m^*$.

Fix a value of $m \in (0, c/r)$.  Define the controlled process $Y^R$ by
$$
dY^R_t = -(r - \la) \, Y^R_t \, dt + {\mu - r \over \sigma} \, Y^R_t \, d \hat B_t + d R_t, \quad Y^R_0 = y > 0,
$$
in which $\hat B$ is a standard Brownian motion with respect to a filtration of a probability space $(\hat \Omega, \hat {\cal F}, \hat \P)$.  Here, $R$ is a right-continuous, non-negative, non-decreasing control that incurs a proportional cost of $m$ when the controller implements it. 

For $y > 0$, define the function $\hat \phi$ by
\begin{equation}\label{eq:hatphi}
\hat \phi(y, m) = \inf_\tau \sup_R \hat \E^y \left[ \int_0^\tau e^{-\la t} \left( c \, Y^R_t \, dt - m \, dR_t \right) + e^{-\la \tau} \left(1 + \al Y^R_\tau \right) \right].
\end{equation}
$\hat \phi$ is the value function for an optimal controller-stopper problem.  Specifically, the controller chooses among processes $R$ in order to maximize the discounted (net) running ``penalty'' to the stopper given by $c \, Y^R_t$ in \eqref{eq:hatphi}, net of the controller's proportional cost $m$.  Then, the stopper chooses the time $\tau$ to stop the game in order to minimize the penalty but has to incur the terminal cost of $1+ \al Y^R_\tau$ when she stops.

\rem{5.7} {The idea of relating the minimum probability of ruin problem and the optimal controller-stopper problem via a duality argument, first appeared in Bayraktar and Young (2011). See, also, Wang and Young (2012a, 2012b) where duality arguments are used for solving related minimum probability of ruin problems. For optimal controller-stopper problems, see, among others, Karatzas and Sudderth (2001), Karatzas and Zamfirescu (2008), Bayraktar et al. (2010) and, more recently, Bayraktar and Huang (2013), Bayraktar and Yao (2014), and Nutz and Zhang (2015). Finally, the controller-stopper problem \eqref{eq:hatphi} is slightly different from the ones appeared in the aforementioned references, and is a so-called ``monotone controller-stopper problem''. For this type of problems, see Karatzas and Shreve (1984) and Bayraktar and Egami (2008).}

Via standard techniques (\O ksendal and Sulem, 2004, Chapter 5), one can show that there exists $\hat{y}_m(m) > 0$ such that the controller implements the control $R$ in order to keep $y \ge \hat{y}_m(m)$.  Specifically, if $Y^R_0 = y < \hat{y}_m(m)$, then the controller immediately moves $Y^R$ to $\hat{y}_m(m)$ and incurs the cost $m (\hat{y}_m(m) - y)$.  Thus, for $y < \hat{y}_m(m)$, we have $\hat \phi(y, m) = -m(\hat{y}_m(m) - y) + \hat \phi(\hat{y}_m(m))$.  After that, the controller exercises instantaneous control to keep $y \ge \hat{y}_m(m)$.

Additionally, one can show that there exists $\hat{y}_{\al m}(m) > \hat{y}_m(m)$ such that the stopper stops the game immediately if $Y^R_0 = y \ge \hat{y}_{\al m}(m)$, and if $y < \hat{y}_{\al m}(m)$, then she stops when $Y^R$ reaches $\hat{y}_{\al m}(m)$.  Thus, if $y \ge \hat{y}_{\al m}(m)$, we have $\hat \phi(y, m) = 1 + \al m y$.  (For later purposes, we make the dependence of $\hat{y}_m$ and $\hat{y}_{\al m}$ upon $m$ explicit by writing $\hat{y}_m(m)$ and $\hat{y}_{\al m}(m)$, respectively.) 

Moreover, $\hat \phi$ is concave with respect to $y$ on $\R^+$ and is the unique classical solution of the following FBP for $y \in [\hat{y}_m(m), \hat{y}_{\al m}(m)]$.
\begin{equation}\label{eq:FBP2}
\left\{
\eqalign{
&\del y^2 f_{yy} - (r - \la) y f_y - \la f + c y = 0,  \cr
&f(\hat{y}_{\al m}(m), m) = 1 + \al m \, \hat{y}_{\al m}(m), \quad f_y(\hat{y}_{\al m}(m), m) = \al m, \cr
&f_y(\hat{y}_m(m), m) = m, \quad f_{yy}(\hat{y}_m(m), m) = 0.}
\right.
\end{equation}
In the following proposition, we present the solution of the FBP \eqref{eq:FBP2}.

\prop{5.8} {Suppose $0 < m < c/r$.  The solution of the free-boundary problem \eqref{eq:FBP2} for $y \in [\hat{y}_m(m), \hat{y}_{\al m}(m)]$ and, hence, the value function of the optimal controller-stopper problem \eqref{eq:hatphi} for such $y$ is given by
\begin{equation}\label{eq:hatphi_sol}
\eqalign{
\hat \phi(y, m) = {c \over r} \, y &- \left[ {B_2 \over B_1 - B_2} + \left( {c \over r} - \al m \right) {1 - B_2 \over B_1 - B_2} \, \hat{y}_{\al m}(m) \right] \left( {y \over \hat{y}_{\al m}(m)} \right)^{B_1}\cr
&+ \left[ {B_1 \over B_1 - B_2} - \left( {c \over r} - \al m \right) {B_1 - 1 \over B_1 - B_2} \, \hat{y}_{\al m}(m) \right] \left( {y \over \hat{y}_{\al m}(m)} \right)^{B_2},
}
\end{equation}
in which $B_1$ and $B_2$ are given by \eqref{eq:B1} and \eqref{eq:B2}, respectively; the free boundary $\hat{y}_m(m) > 0$ is given by
$$
\hat{y}_m(m) = {\hat{y}_{\al m}(m) \over x(m)},
$$
in which $x(m) > 1$ is given in Lemma 5.2; and the free boundary $\hat{y}_{\al m}(m) > \hat{y}_m(m)$ is given in terms of $x(m)$ by
\begin{equation}\label{eq:yhat_alm}
\eqalign{
{1 \over \hat{y}_{\al m}(m)} &= \left({c \over r} - \al m \right) \cr
&\quad - \left( {c \over r} - m \right) \left[ {1 - B_2 \over B_1 (B_1 - B_2)} \, x(m)^{B_1 - 1} + {B_1 - 1 \over B_2 (B_1 - B_2)} \, x(m)^{B_2 - 1} \right].
}
\end{equation}
Moreover, $\hat \phi(\cdot,m)$ is ${\cal C}^2$ and is increasing and concave on $[\hat{y}_m(m), \hat{y}_{\al m}(m)]$.}

\pf It is easy to show that the expression in \eqref{eq:hatphi_sol} satisfies the differential equation in \eqref{eq:FBP2} and that it satisfies the free-boundary conditions $\hat \phi_y(\hat{y}_m(m), m) = m$ and $\hat \phi_{yy}(\hat{y}_m(m), m) = 0$.  The expression for $x(m)$ in \eqref{eq:xm} implies that $\hat \phi$ in \eqref{eq:hatphi_sol} satisfies the free-boundary condition $\hat \phi_y(\hat{y}_{\al m}(m), m) = \al m$; similarly, the expression in \eqref{eq:yhat_alm} implies $\hat \phi(\hat{y}_{\al m}(m), m) = 1 + \al m \, \hat{y}_{\al m}(m)$.

Finally, we show that $\hat \phi$ given in \eqref{eq:hatphi_sol} is, indeed, increasing and concave with respect to $y$ on $[\hat{y}_m(m), \hat{y}_{\al m}(m)]$, as expected, because $\hat \phi$ defined in \eqref{eq:hatphi} uniquely solves \eqref{eq:FBP2}.  To that end, it follows from \eqref{eq:hatphi_sol}, \eqref{eq:yhat_alm}, and \eqref{eq:xm} that
\begin{equation}\label{eq:phihaty}
\hat \phi_y(y, m) = {c \over r}  - \left( {c \over r} - m \right) \left[  {1 - B_2 \over B_1 - B_2} \left( {y \over \hat{y}_m(m)} \right)^{B_1 - 1} + {B_1 - 1 \over B_1 - B_2} \left( {y \over \hat{y}_m(m)} \right)^{B_2 - 1} \right],
\end{equation}
and
\begin{equation}\label{eq:phihatyy}
\hat \phi_{yy}(y, m) = - \left( {c \over r} - m \right)  {(B_1 - 1)(1 - B_2) \over (B_1 - B_2) \hat{y}_m(m)} \left[  \left( {y \over \hat{y}_m(m)} \right)^{B_1 - 2} - \left( {y \over \hat{y}_m(m)} \right)^{B_2 - 2} \right].
\end{equation}
Because $\hat \phi_{yy} < 0$ for $\hat{y}_m(m) < y \le \hat{y}_{\al m}(m)$ and because $\hat \phi_y(\hat{y}_{\al m}(m), m) = \al m > 0$, it follows that $\hat \phi$ is increasing and concave with respect to $y$ on $[\hat{y}_m(m), \hat{y}_{\al m}(m)]$.  \qed

Because $\hat \phi$ is concave with respect to $y$, we can define its convex dual $\Phi$, via the Legendre transform, by
\begin{equation}\label{eq:Phi}
\Phi(w, m) = \max_y \left( \hat \phi(y, m) - wy \right).
\end{equation}
In the next proposition, we show that $\Phi$ is a probability of drawdown under a restriction on the admissible investment strategies.

\prop{5.9} {$\Phi$ in $\eqref{eq:Phi}$ is the minimum probability of lifetime drawdown on $\{(w,m) \in ({\bf R}^+)^2: \al m \le w \le m, \, 0 < m < c/r \}$ under the restriction that $M_t = m$ almost surely, for all $t \ge 0$, that is, wealth may not grow larger than $m$.}

\pf  As in the proof of Proposition 5.5, one can show that $\Phi$ is the classical solution of the following boundary-value problem.
%
\begin{equation}\label{eq:BVP2}
\left\{
\eqalign{
&\la h = (r w - c) h_w + \min_\pi \left[ (\mu - r) \pi h_w + {1 \over 2} \sigma^2 \pi^2 h_{ww} \right], \cr
&h(\al m, m) = 1, \quad \lim_{w \to m-} {h_w(w, m) \over h_{ww}(w, m)} = 0.}
\right.
\end{equation}
Note that the condition $\lim_{w \to m-} {h_w(w, m) \over h_{ww}(w, m)} = 0$ is equivalent to $M_t = m$ almost surely, for all $t \ge 0$.  Indeed, the optimal investment in the risky asset is given by
$$
\pi^*_t = - \, {\mu - r \over \sigma^2} \, {\Phi_w(W^*_t, m) \over \Phi_{ww}(W^*_t, m)},
$$
in which $W^*$ is the optimally controlled wealth.  Because $\pi^*_t = 0$ almost surely when wealth reaches $m$ and because the consumption rate $c$ is greater than $rm$, wealth can never become larger than $m$.

From a verification result similar to Corollary 3.2, we deduce that $\Phi$ is the minimum probability of lifetime drawdown under the restriction that wealth cannot grow larger than the current maximum $m$.  \qed

We now present the main result of the paper regarding the optimal probability of drawdown.  We have already identified the optimal probability of drawdown for $m > c/r$ and $m^* \le m \le c/r$ in Theorems 4.1 and 5.6, respectively.  The following theorem completes the picture by showing that $\Phi$ defined in \eqref{eq:Phi} is the (unrestricted) minimum probability of lifetime drawdown $\phi$ when $0 < m \le m^*$.

\th{5.10} {Assume that the conditions of Proposition 5.4 hold such that $z(m)$ is the solution of ODE \eqref{eq:ODE1} on $[m^*,c/r]$ satisfying $z(c/r) = 0$, and let $\Psi$ and $\Phi$ be given by \eqref{eq:Psi} and \eqref{eq:Phi}, respectively. Then, the function $\phi$ on $\D = \{(w,m) \in ({\bf R}^+)^2: \al m \le w \le m, \, 0 < m < c/r \}$ defined by
\begin{equation}\label{eq:varphi}
	\phi(w,m) = 
	\left\{
	\eqalign{
		\Phi(w,m),	&\quad 0 < m < m^*,\cr
		\Psi(w,m),	&\quad m^*\le m < c/r,
	}
	\right.
\end{equation}
is the $($unrestricted$)$ minimum probability of drawdown on $\D$.}

\pf See Appendix C.\qed

Figure 5 illustrates the optimal probability of drawdown \eqref{eq:varphi} for $0 < m < c/r$. In particular, note that $\phi$ is smooth except at $m = m^*$, where it is not differentiable with respect to $m$.

\begin{figure}[b]
\hspace{-5em}
\adjustbox{trim={0.015\width} {0.0\height} {0.12\width} {0.06\height},clip}{\includegraphics[scale=0.35,page=1]{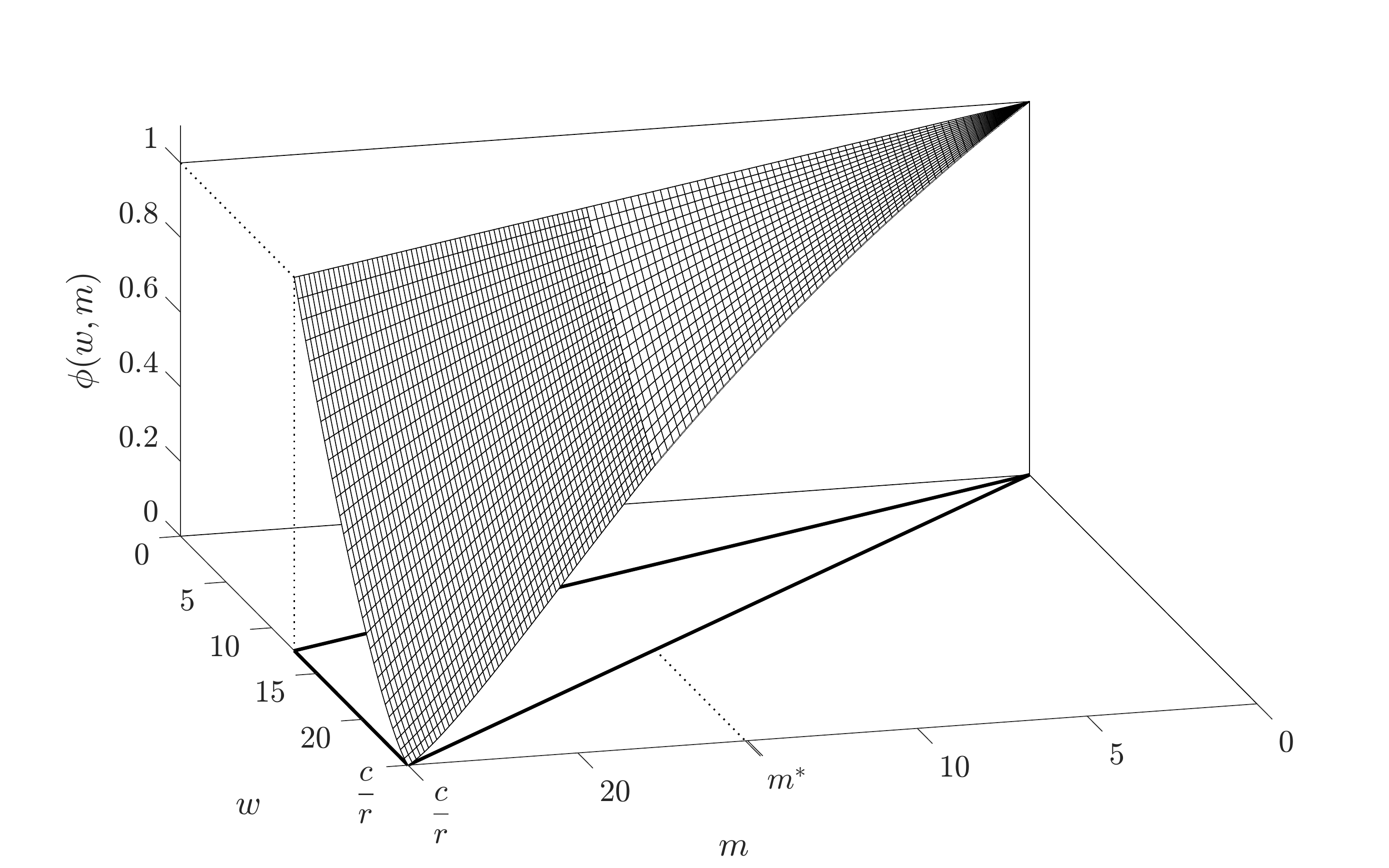}}
\hspace{1em}
\adjustbox{trim={0.015\width} {0.0\height} {0.05\width} {0.06\height},clip}{\includegraphics[scale=0.35,page=1]{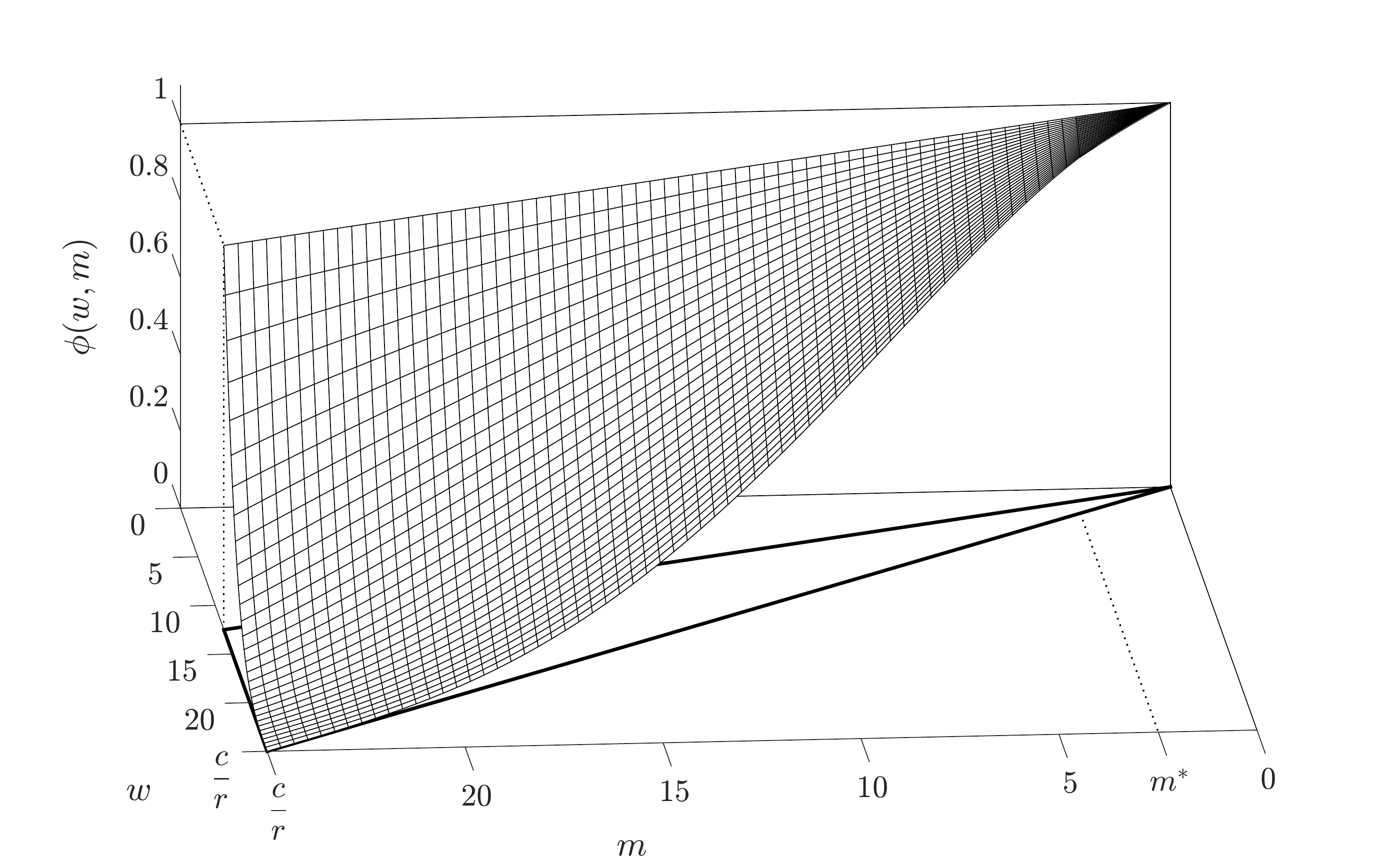}}
\nobreak
\centerline{{\bf Figure 5}: Optimal probability of drawdown for $\al\, m \le w \le m$ and $0 < m < c/r$.}\vspace{1em}
\end{figure}

\rem{5.11} Theorem 5.10 tells us that if the so-called initial maximum wealth $m$ is low enough, specifically $m \le m^*$, then to minimize the probability of lifetime drawdown, the individual will not allow her wealth to exceed the current maximum $m$.  \qed


\rem{5.12} {One can combine the results of Sections 5.1 and 5.2 (semi-)explicitly. Assume that the conditions of Proposition 5.4 hold such that $z(m)$ is the unique solution of ODE \eqref{eq:ODE1} on $[m^*,c/r]$ satisfying $z(c/r)=0$. Define the function $\eta:[0, c/r]\to[0,1]$ by
\begin{equation}\label{eq:eta}
	\eta(m) = 
	\left\{
	\eqalign{
		1/x(m),	&\quad 0 \le m \le m^*,\cr
		z(m), 	&\quad m^* \le m \le c/r,}
	\right.
\end{equation}
in which $x(m)$ is given by \eqref{eq:xm}. Furthermore, define $y_{\al m}(m)$ in terms of $\eta(m)$ by
\begin{equation}\label{eq:yalm_eta}
\eqalign{
& {1 \over y_{\al m}(m)} \, {B_1 B_2 \over B_1 - B_2} \left( \eta(m)^{B_1-1} - \eta(m)^{B_2-1} \right)  \cr
& \quad = \left( {c \over r} - m \right) - \left( {c \over r} - \al m \right) \left[ {B_1 (1 - B_2) \over B_1 - B_2} \eta(m)^{B_1-1} + {B_2(B_1 - 1) \over B_1 - B_2} \eta(m)^{B_2-1} \right],}
\end{equation}
and $y_m(m)=\eta(m) \, y_{\al m}(m)$. Then, for $\al m \le w \le m$ and $0 \le m \le c/r$, the minimum probability of lifetime drawdown $\phi$ is given by
\begin{equation}\label{eq:phi}
\eqalign{
\phi(w, m) &= {B_1 - 1 \over B_1 - B_2} \left[ B_2 + \left( {c \over r} - \al m \right) (1 - B_2) \, y_{\al m}(m) \right] \left( {y \over y_{\al m}(m)} \right)^{B_1} \cr
& \quad + {1 - B_2 \over B_1 - B_2} \left[ B_1 - \left( {c \over r} - \al m \right) (B_1 - 1) \, y_{\al m}(m) \right] \left( {y \over y_{\al m}(m)} \right)^{B_2},
}
\end{equation}
in which $y \in [y_m(m), y_{\al m}(m)]$ uniquely solves
\begin{equation}\label{eq:I}
\eqalign{
{c \over r} - w &=  {B_1  \over B_1 - B_2} \left[ { B_2 \over  y_{\al m}(m)} + \left( {c \over r} - \al m \right) (1 - B_2) \right] \left( {y \over y_{\al m}(m)} \right)^{B_1 - 1} \cr
& \quad - {B_2 \over B_1 - B_2} \left[ {B_1 \over  y_{\al m}(m)} - \left( {c \over r} - \al m \right) (B_1 - 1) \right] \left( {y \over y_{\al m}(m)} \right)^{B_2 - 1}.
}
\end{equation}
For wealth between $\al m$ and $m$, the corresponding optimal investment strategy $\pi^*$ is given in feedback form by $\pi^*_t = \pi^*(W^*_t, M^*_t)$, in which
\begin{equation}\label{eq:piStar}
\eqalign{
\pi^*(w, m) &= {\mu - r \over \sigma^2} \,  {B_1 (B_1 - 1) \over B_1 - B_2} \left[  {B_2 \over y_{\al m}(m)} +  \left( {c \over r} - \al m \right) (1 - B_2)  \right] \left( {y \over y_{\al m}(m)} \right)^{B_1 - 1} \cr
& \quad +  {\mu - r \over \sigma^2} \,  {B_2 (1 - B_2) \over B_1 - B_2} \left[  {B_1 \over y_{\al m}(m)} -  \left( {c \over r} - \al m \right) (B_1 - 1)  \right]  \left( {y \over y_{\al m}(m)} \right)^{B_2 - 1}.}
\end{equation}
Finally, for $0 < m \le m^*$, $\pi^*_t$ is such that $M_t = m$ almost surely, for all $t \ge 0$; by contrast, for $m^*< m < c/r$, $\pi^*_t$ allows $M_t$ to increase. \qed}


\subsect{5.3  Properties of the optimal investment strategy for $0 < m \le m^*$}

When $0 < m \le m^*$, one can write the investment strategy \eqref{eq:piStar} more simply as follows:
\begin{equation}\label{eq:piStarhat}
\pi^*(w, m) = {\mu - r \over \sigma^2} \left( {c \over r} - m \right) {(B_1 - 1)(1 - B_2) \over B_1 - B_2} \left[ \left( {y \over \hat y_m(m)} \right)^{B_1 - 1} - \left( {y \over \hat y_m(m)} \right)^{B_2 - 1} \right].
\end{equation}
Indeed, $\pi^*(w, m) = -{\mu - r \over \sigma^2} \, y \, \hat \phi_{yy}$, and the expression for $\hat \phi_{yy}$ in \eqref{eq:phihatyy} gives us \eqref{eq:piStarhat}.  In the next four propositions, we study properties of the optimal investment strategy given in \eqref{eq:piStarhat}.

From Young (2004), we know that the optimal amount invested in the risky asset to minimize the probability of lifetime ruin is given by the expression in (4.1).  When $m < c/r$, this investment strategy allows the maximum wealth to increase beyond $m$; thus, this investment strategy is not the optimal one corresponding to the drawdown problem when $0 < m \le m^*$.

Note that the investment strategy given by (4.1) decreases as wealth $w$ increases.  The same is true for the optimal investment strategy when  $0 < m \le m^*$, as we demonstrate in the following proposition.

\prop{5.13} {For $0 < m \le m^*$, the optimal amount invested in the risky asset decreases with respect to $w \in (\al m, m)$, and $\pi^*(m, m) = 0$.}

\pf  From $y = - \phi_w$, it follows that ${\partial y \over \partial w} = - \phi_{ww} < 0$ for $w \in (\al m, m)$.  Thus, the optimal amount invested in the risky asset decreases with respect to $w$ if and only if the expression for $\pi^*$ in \eqref{eq:piStarhat} increases with respect to $y$, which is equivalent to
$$
(B_1 - 1) \left( {y \over y_m(m)} \right)^{B_1 - B_2} + (1 - B_2),
$$
which is clearly positive.  Furthermore, if $w = m$, then $y = y_m(m)$, and it follows from \eqref{eq:piStarhat} that $\pi^*(m, m) = 0$.  \qed

Because the investment strategy in (4.1) allows wealth to increase above $m$ and the investment strategy in \eqref{eq:piStarhat} does not, we expect the former to be larger than the latter, which we show in the next proposition.

\prop{5.14} {For  $\alpha m \le w \le m$ and $0 < m \le m^*$, the optimal amount invested in the risky asset satisfies
$$
\pi^*(w, m) < {\mu - r \over \sigma^2} \, {1 \over \gam - 1} \, \left( {c \over r} -  w \right).
$$}

\pf  Use \eqref{eq:piStarhat} and \eqref{eq:phihaty} to substitute for $\pi^*$ and ${c \over r} - w = {c \over r} - \hat \phi_y$, respectively, in the above inequality.  Simplify to learn that it is equivalent to $B_2 < B_1$, which is true because $B_2$ is negative and $B_1$ is positive.   \qed

For the drawdown level $\al m$ small enough, the expression $ {\mu - r \over \sigma^2} \, {1 \over \gam - 1} \, \left( {c \over r} -  w \right)$ is greater than $w$ when wealth is close to $\al m$; therefore, in minimizing the probability of ruin, the individual will leverage her wealth in order to avoid ruin.  However, because $\pi^*(w, m)$ is less than this, leveraging will be less under the goal of minimizing the probability of drawdown than when minimizing the probability of lifetime ruin when $0 < m \le m^*$.  This decreased leveraging is prudent because it is unlikely that a financial advisor will recommend that an individual invest more in the risky asset than her current wealth.

Because the difference ${\mu - r \over \sigma^2} \, {1 \over \gam - 1} \, \left( {c \over r} -  w \right) - \pi^*(w, m)$ is positive for  $\alpha m \le w \le m$ and $0 < m \le m^*$, we ask if the difference in the investment strategies is monotone with respect to $w$.  The next proposition proves that this is the case.

\prop{5.15} {For $0 < m \le m^*,$ the difference ${\mu - r \over \sigma^2} \, {1 \over \gam - 1} \, \left( {c \over r} -  w \right) - \pi^*(w, m)$ increases with respect to $w \in (\al m, m)$.}

\pf  In terms of $y$, this difference equals
$$
{\mu - r \over \sigma^2} \left( {c \over r} - m \right) (B_1 - 1) \left( {y \over y_m(m)} \right)^{B_2 - 1},
$$
which is clearly decreasing with respect to $y$ because $B_2 - 1 < 0$.  Thus, this difference is increasing with respect to $w$ because ${\partial y \over \partial w} = - \phi_{ww} < 0$.  \qed

Recall that when $0 < m \le m^*$, investment in the risky asset is positive for wealth strictly less than $m$, and it decreases to 0 as wealth increases to $m$.  Thus, for a given value of $w$, we anticipate that the optimal amount invested in the risky asset increases with respect to $m$ because as $m$ increases, the point at which the investment equals 0 increases.  The following proposition shows that our intuition is correct.

\prop{5.16} {For  $\alpha m \le w \le m$ and $0 < m < m^*,$ the optimal amount invested in the risky asset increases with respect to $m$.}

\pf From \eqref{eq:piStarhat}, we see that $\pi^*$ depends on $m$ via the ratio ${y \over y_m}$, whose dependence on $m$ is given in ${c \over r} - w = \hat \phi_y$ with $\hat \phi_y$ given in \eqref{eq:phihaty}.  If we fully differentiate \eqref{eq:phihaty} with respect $m$, we obtain
$$
\eqalign{
&\left[ {1 - B_2 \over B_1 - B_2} \left( {y \over y_m} \right)^{B_1 - 1} + {B_1 - 1 \over B_1 - B_2} \left( {y \over y_m} \right)^{B_2 - 1} \right] \cr
&\quad = \left( {c \over r} - m \right) {(B_1 - 1)(1 - B_2) \over B_1 - B_2} \left[ \left( {y \over y_m} \right)^{B_1 - 2} - \left( {y \over y_m} \right)^{B_2 - 2} \right] {\partial \over \partial m} {y \over y_m}.
}
$$
Thus, by writing $v = {y \over y_m}$, we obtain
\begin{align}
	\frac{\partial \pi^*}{\partial m} &\propto \frac{\partial}{\partial m} \left\{ \left( \frac{c}{r} - m \right) \left[ \left( \frac{y}{y_m} \right)^{B_1 - 1} - \left( \frac{y}{y_m} \right)^{B_2 - 1} \right] \right\} \nonumber\\
	&= - \left( v^{B_1 - 1} - v^{B_2 - 1} \right) + \left( \frac{c}{r} - m \right) \big( (B_1 - 1) v^{B_1 - 2} + (1 - B_2) v^{B_2 - 2} \big) \frac{\partial}{\partial m} \frac{y}{y_m} \nonumber\\
	&\propto - \frac{(B_1 - 1)(1 - B_2)}{B_1 - B_2}  \left( v^{B_1 - 1} - v^{B_2 - 1} \right)^2 \nonumber\\
	& \quad + \big( (B_1 - 1) v^{B_1 - 1} + (1 - B_2) v^{B_2 - 1} \big) \left( \frac{1 - B_2}{B_1 - B_2} v^{B_1 - 1} + \frac{B_1 - 1}{B_1 - B_2} v^{B_2 - 1} \right) \nonumber\\
	&= (B_1 - B_2) \, v^{B_1 - 1} v^{B_2 - 2} > 0.\tag*{$\square$}
\end{align}

We end this section with a brief discussion on the properties of the optimal allocation for $m^*<m<c/r$. Here, in contrast to the case $0<m\le m^*$, we do not have a simplified expression for the optimal allocation akin to \eqref{eq:piStarhat}. This is mainly because of the lack of an expression for the solution of ODE \eqref{eq:ODE1}. Therefore, proving the properties of the optimal allocation directly from its expression in \eqref{eq:piStar} is cumbersome. Instead, we opt to illustrate the properties numerically and invite the interested reader to prove them.

Figure 6 suggests that, for any fixed value of $m\in(0,c/r)$, the optimal allocation $\pi^*(w,m)$ decreases as the wealth $w$ increases. Note, also, that $\pi^*(m,m)> 0$ for $m^*<m<c/r$ (which has already been illustrated in Figure 4). Thus, we conjecture that the first part of Proposition 5.13 holds for $m^*<m<c/r$.

Figure 7 illustrates that for fixed values of $w\in(0,c/r)$, the optimal allocation increases as the high-water mark $m$ increases, thus, suggesting that Proposition 5.16 also holds for $m^*<m<c/r$.

Finally, Figure 8 suggests that the result of Proportions 5.16 also holds in the case $m^*<m<c/r$, by showing that the values of
\[
	\min_{w\in[\alpha\,m,m]}\Big\{ {\mu - r \over \sigma^2} \, {1 \over \gamma - 1} \, \left( {c \over r} -  w \right) - \pi^*(w,m)\Big\}
\]
are positive for any $0<m<c/r$.

\begin{figure}[t]
\hspace{-5em}
\adjustbox{trim={0.015\width} {0.0\height} {0.12\width} {0.06\height},clip}{\includegraphics[scale=0.33,page=1]{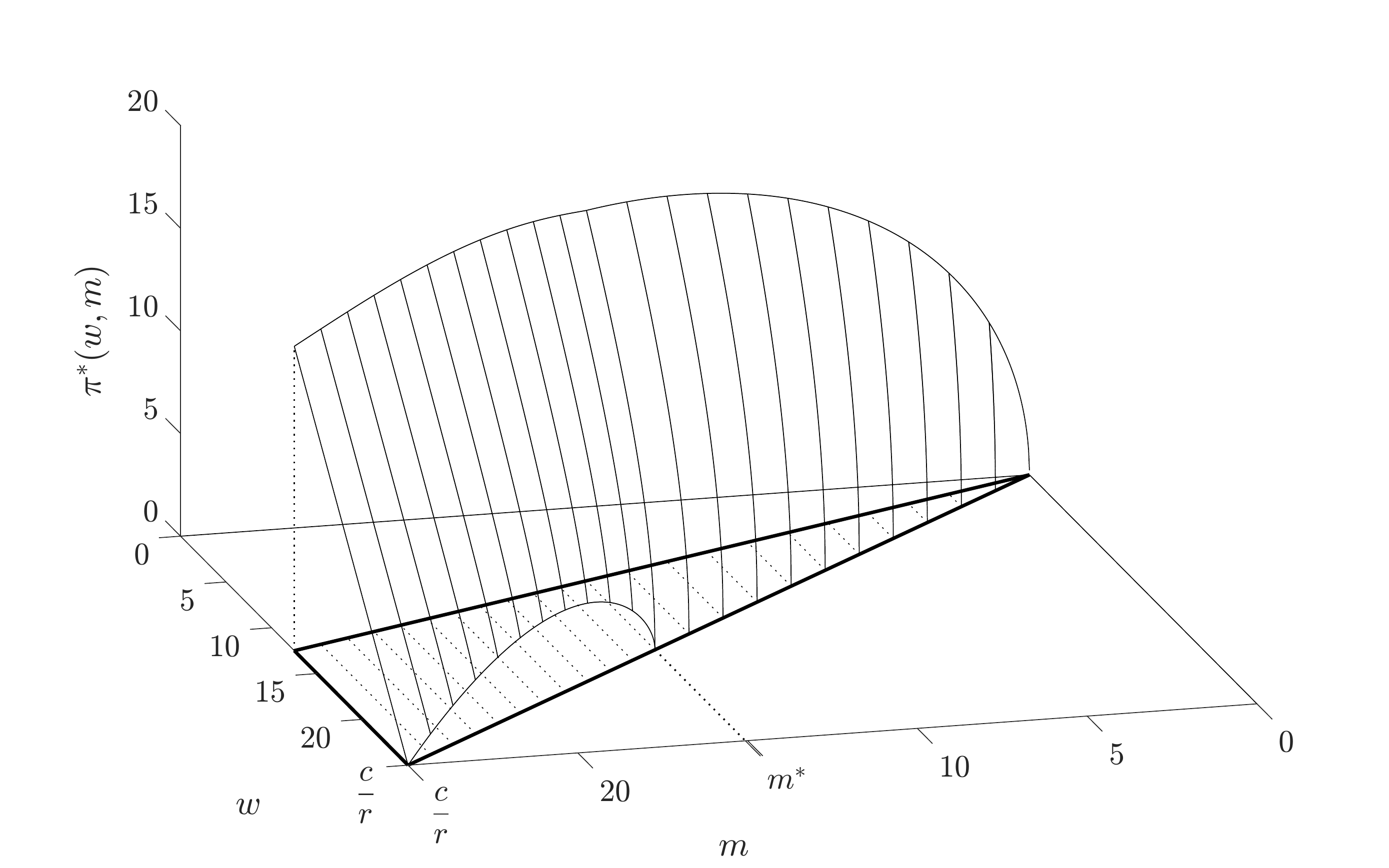}}
\hspace{1em}
\adjustbox{trim={0.015\width} {0.0\height} {0.05\width} {0.06\height},clip}{\includegraphics[scale=0.33,page=1]{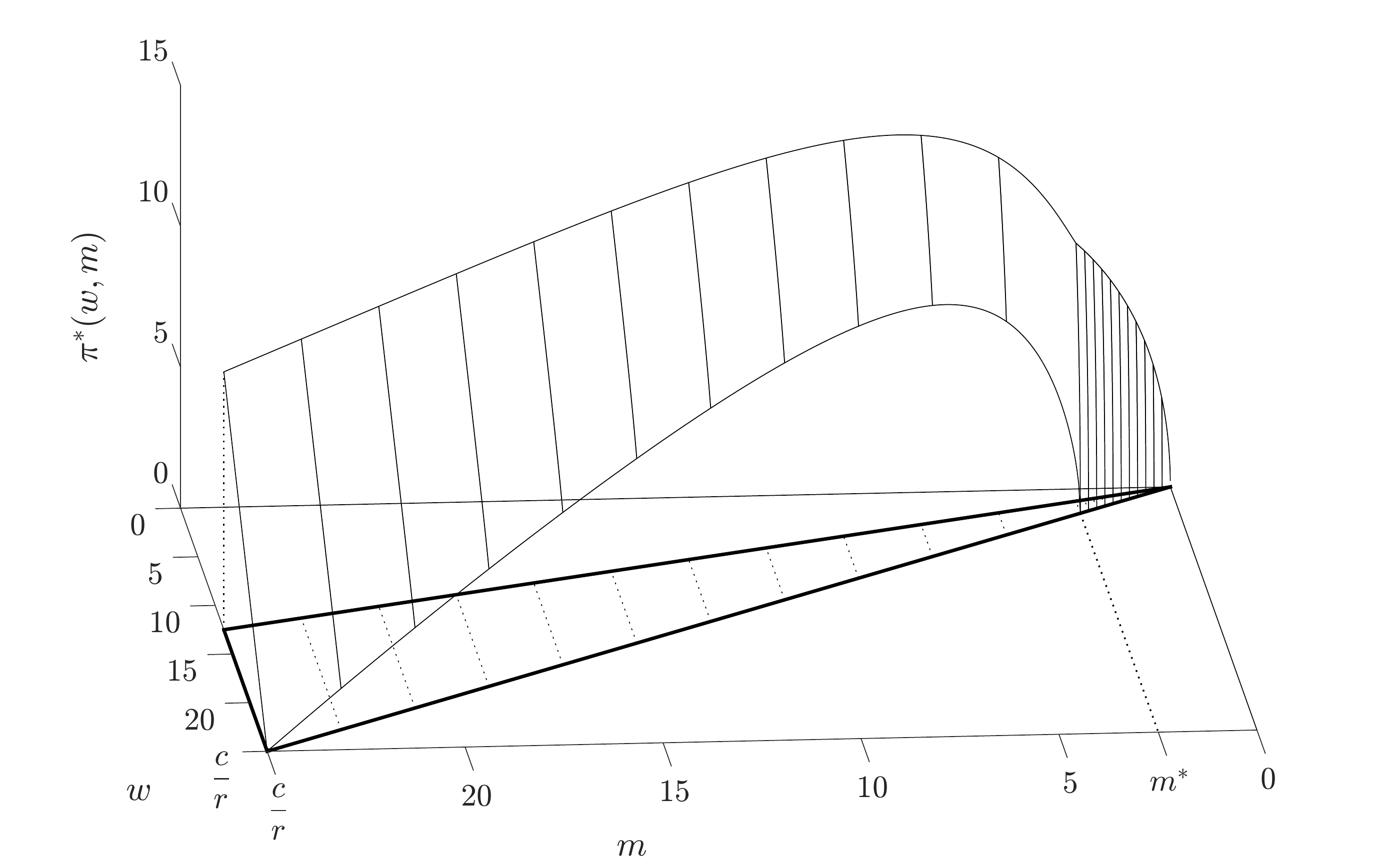}}
\nobreak
\centerline{{\bf Figure 6}: The change in the optimal allocation when $m$ is fixed and $w$ is changing.}\vspace{1em}
\end{figure}

\begin{figure}[H]
\hspace{-5em}
\adjustbox{trim={0.015\width} {0.0\height} {0.12\width} {0.06\height},clip}{\includegraphics[scale=0.33,page=1]{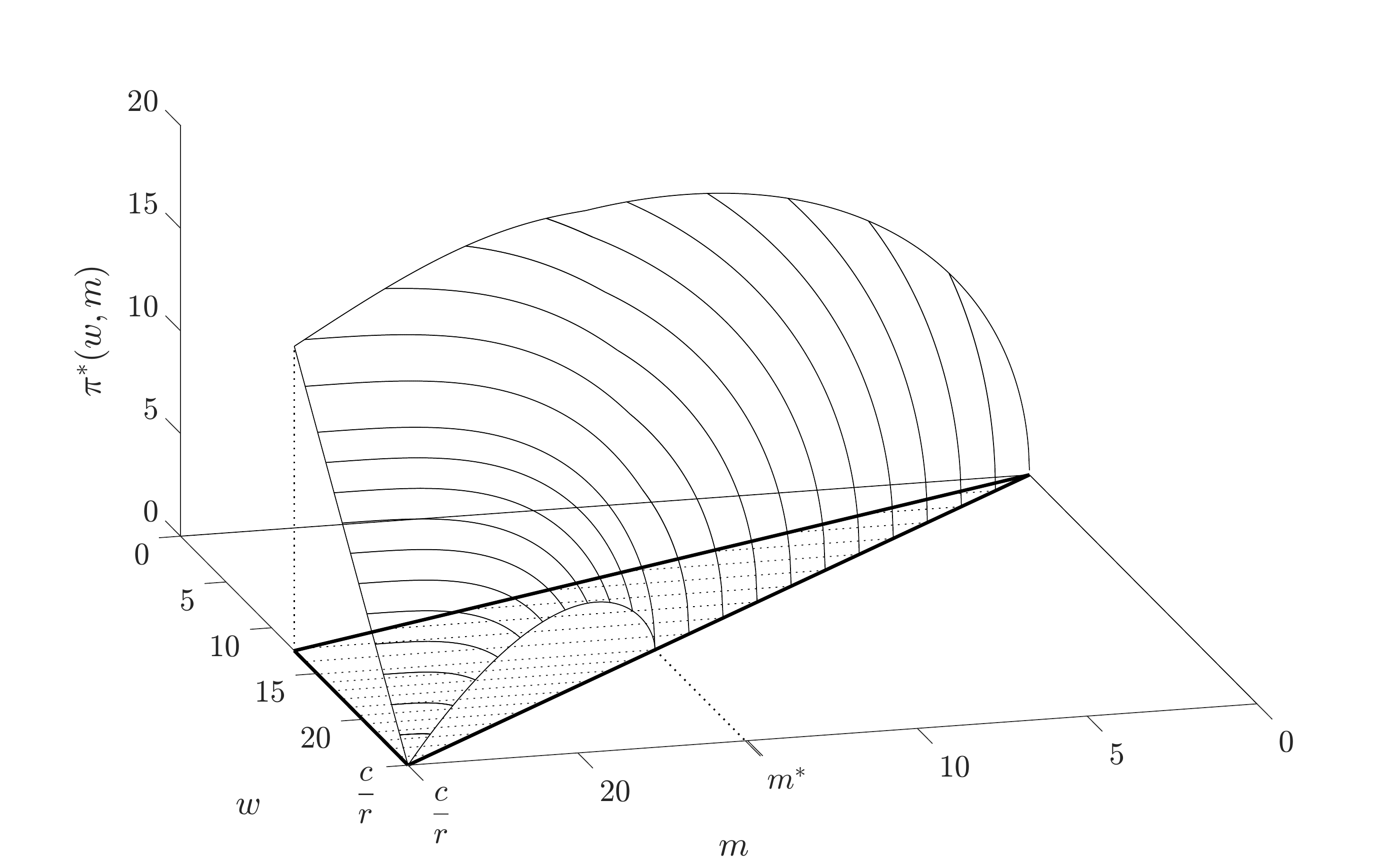}}
\hspace{1em}
\adjustbox{trim={0.015\width} {0.0\height} {0.05\width} {0.06\height},clip}{\includegraphics[scale=0.33,page=1]{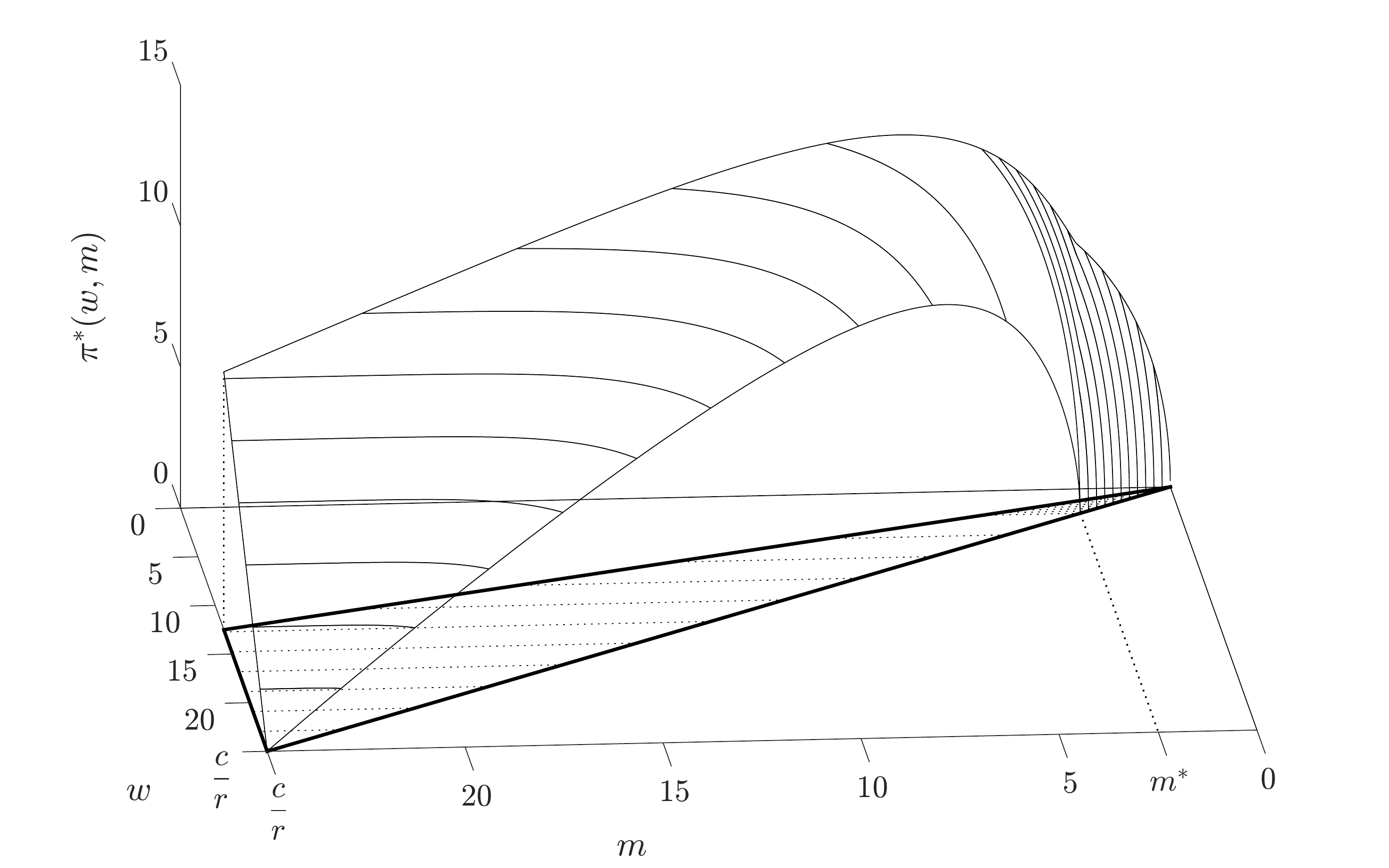}}
\nobreak
\centerline{{\bf Figure 7}: The change in the optimal allocation when $w$ is fixed and $m$ is changing.}\vspace{1em}
\end{figure}

\begin{figure}[H]
\hspace{-5em}
\adjustbox{trim={0.015\width} {0.0\height} {0.12\width} {0.06\height},clip}{\includegraphics[scale=0.33,page=1]{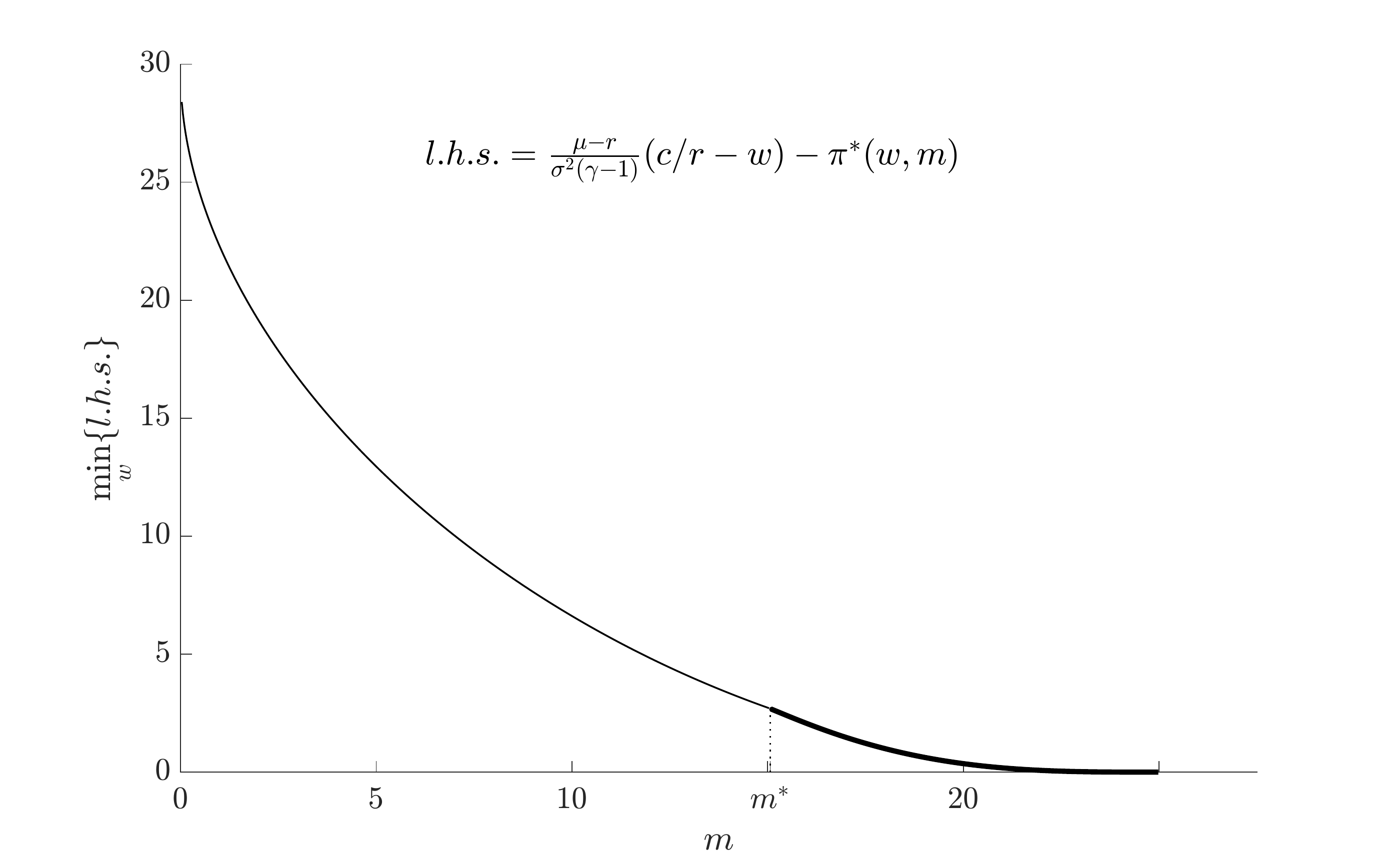}}
\hspace{1em}
\adjustbox{trim={0.015\width} {0.0\height} {0.05\width} {0.06\height},clip}{\includegraphics[scale=0.33,page=1]{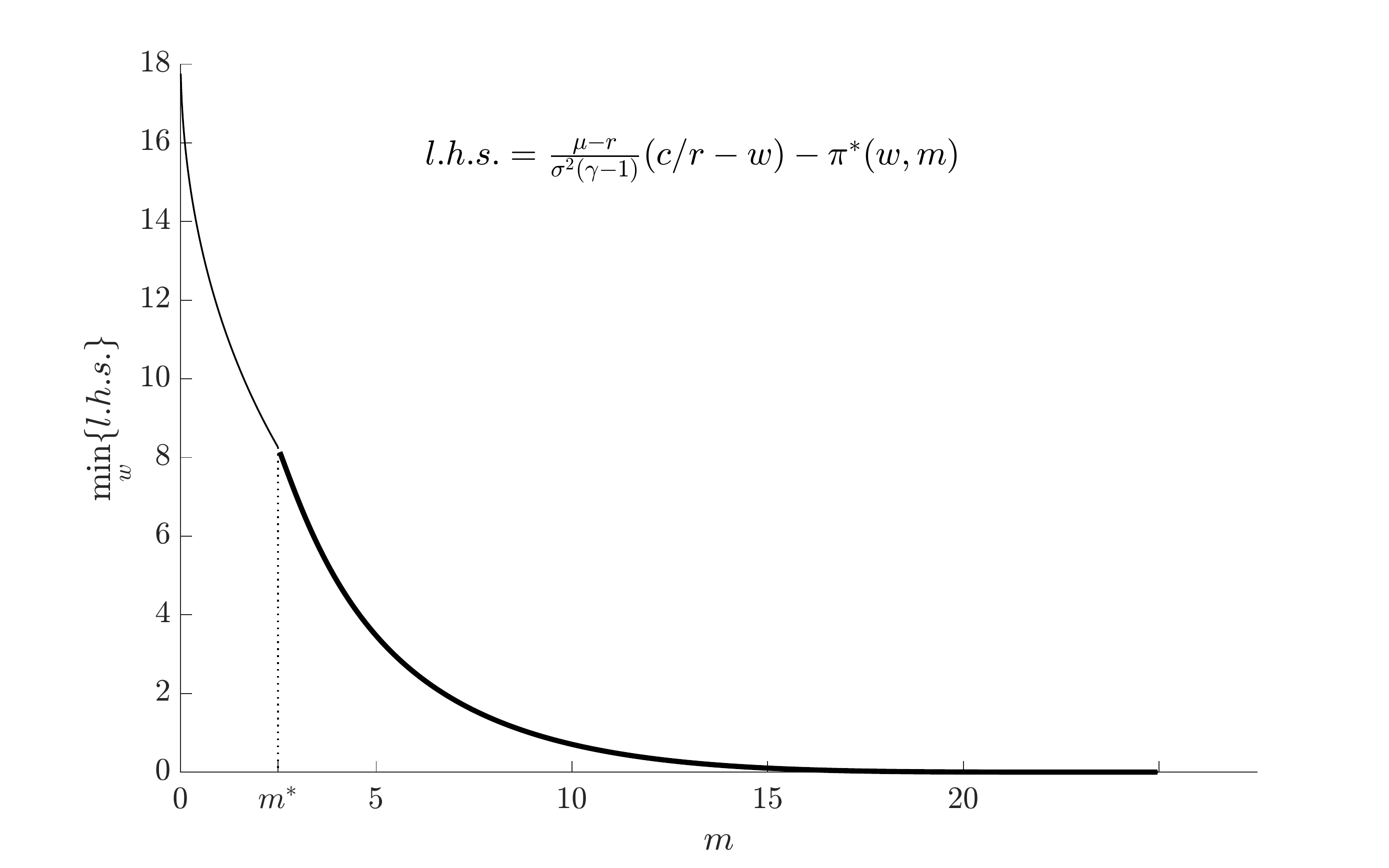}}
\nobreak
\centerline{{\bf Figure 8}: Verifying the inequality $\pi^*(w, m) < {\mu - r \over \sigma^2} \, {1 \over \gamma - 1} \, \left( {c \over r} -  w \right)$ for $0<m<c/r$ and $\al m\le w\le m$.}\vspace{1em}
\end{figure}

\sect{6. Summary and Conclusions}

In this paper, we found the optimal investment strategy to minimize the probability that an individual's wealth drops to a given proportion of maximum wealth before she dies, that is, the individual wishes to minimize the probability of lifetime drawdown.  We assumed that the individual consumes at a constant rate $c$, and the safe level for this problem ${c \over r}$ is identical to the safe level for minimizing the probability of lifetime ruin.  In Section 5.2, we showed that the minimum probability of drawdown when maximum wealth does not increase is the Legendre dual of the value function of an optimal controller-stopper problem.  In minimizing the probability of lifetime ruin with random consumption, Bayraktar and Young (2011) found a similar relationship. 

We learned the following about the optimal investment strategy when minimizing the probability of lifetime drawdown while consuming at a constant rate.

\smallskip

\item{$\bullet$} If $\al m < w < {c \over r} \le m$, then the optimal investment strategy is {\it identical} to the strategy for minimizing the probability of lifetime ruin.

\item{$\bullet$} If $\al m < w \le m \le m^* < {c \over r}$, then the optimal investment strategy is such that maximum wealth never increases above the current maximum $m$.  Intuitively, if the individual were to allow maximum wealth to increase, then the drawdown level of $\al$ times the new maximum would be too great given the {\it constant} rate of consumption.

\item{$\bullet$} If $\al m < w < m$ and $m^* < m < {c \over r}$, then the optimal investment strategy allows maximum wealth to increase to ${c \over r}$.  Intuitively, the individual wishes to increase her wealth in order to fund her consumption.

\medskip

In general, there is a trade-off in allowing maximum wealth to increase.  On the one hand, the drawdown level increases, which {\it could} make drawdown more likely; on the other hand, wealth increases, which helps fund the constant rate of consumption and {\it could} make drawdown less likely.  For $m < m^*$, the former is the case; for $m^* < m < {c \over r}$, the latter is the case.


\bigskip

\centerline{\bf Acknowledgments}

\medskip

The second author thanks the National Science Foundation for financial support under grant number DMS-0955463.  The third author thanks the Cecil J. and Ethel M. Nesbitt Professorship for financial support.

\sect{References}

\noindent \hangindent 20 pt  Angoshtari, Bahman, Erhan Bayraktar, and Virginia R. Young. Optimal Investment to Minimize the Probability of Drawdown. To appear in {\it Stochastics}. (2016)

\smallskip \noindent \hangindent 20 pt Bayraktar, Erhan, and Masahiko Egami. 2008. {\it An analysis of monotone follower problems for diffusion processes.} {\it Mathematics of Operations Research}, 33(2): 336-350.

\smallskip \noindent \hangindent 20 pt Bayraktar, Erhan, and Yu-Jui Huang. 2013. {\it On the multidimensional controller-and-stopper games.} {\it SIAM Journal on Control and Optimization}, 51(2): 1263-1297.

\smallskip \noindent \hangindent 20 pt Bayraktar, Erhan, Ioannis Karatzas, and Song Yao. 2010. {\it Optimal stopping for dynamic convex risk measures.} {\it Illinois Journal of Mathematics}, 54(3): 1025-1067.

\smallskip \noindent \hangindent 20 pt Bayraktar, Erhan, and Song Yao. 2014 {\it On the robust optimal stopping problem.} {\it SIAM Journal on Control and Optimization}, 52(5): 3135-3175.

\smallskip \noindent \hangindent 20 pt  Bayraktar, Erhan, and Virginia R. Young. 2007. Correspondence Between Lifetime Minimum Wealth and Utility of Consumption. {\it Finance and Stochastics} ,11(2): 213-236.

\smallskip \noindent \hangindent 20 pt  Bayraktar, Erhan and Virginia R. Young. 2011.  Proving the Regularity of the Minimal Probability of Ruin via a Game of Stopping and Control.  {\it Finance and Stochastics}, 15: 785-818.

\smallskip \noindent \hangindent 20 pt Chen, Xinfu, David Landriault, Bin Li, and Dongchen Li. 2015. On Minimizing Drawdown Risks of Lifetime Investments. Working Paper. University of Waterloo.

\smallskip \noindent \hangindent 20 pt Cvitani\'c, Jaksa, and Ioannis Karatzas. 1995. On Portfolio Optimization Under Drawdown Constraints. {\it IMA Lecture Notes in Mathematical Applications}, 65: 77-78.


\smallskip \noindent \hangindent 20 pt Grossman, Sanford J. and Zhongquan Zhou. 1993. Optimal investment Strategies for Controlling Drawdowns. {\it Mathematical Finance}, 3(3): 241-276.

\smallskip \noindent \hangindent 20 pt Karatzas, Ioannis, and Steven E. Shreve. (1984) {\it Connections between optimal stopping and singular stochastic control I. Monotone follower problems.} {\it SIAM Journal on Control and Optimization}, 22(6): 856-877.

\smallskip \noindent \hangindent 20 pt Karatzas, Ioannis, and William D. Sudderth. 2001. {\it The controller-and-stopper game for a linear diffusion.} {\it Annals of Probability}, 29(3): 1111-1127.

\smallskip \noindent \hangindent 20 pt Karatzas, Ioannis, and Ingrid-Mona Zamfirescu. 2008. {\it Martingale approach to stochastic differential games of control and stopping.} {\it The Annals of Probability}, 36(4): 1495-1527.

\smallskip \noindent \hangindent 20 pt Kardaras, Constantinos, Jan Ob\l \'oj, and Eckhard Platen. 2014. The Num\'eraire Property and Long-Term Growth Optimality for Drawdown-Constrained Investments. {\it Mathematical Finance}. doi: 10.1111/mafi.12081

\smallskip \noindent \hangindent 20 pt Moore, Kristen S., and Virginia R. Young. 2006. Optimal and Simple, Nearly-Optimal Rules for Minimizing the Probability of Financial Ruin in Retirement. {\it North American Actuarial Journal}, 10(4): 145-161.

\smallskip \noindent \hangindent 20 pt Nutz, Marcel, and Jianfeng Zhang. 2015. {\it Optimal stopping under adverse nonlinear expectation and related games.} {\it The Annals of Applied Probability} 25(5): 2503-2534.

\smallskip \noindent \hangindent 20 pt \O ksendal, Bernt, and Agnes Sulem. 2004. {\it Applied Stochastic Control of Jump Diffusions}. New York: Springer.

\smallskip \noindent \hangindent 20 pt Walter, Wolfgang. {\it Differential and Integral Inequalities.} Springer-Verlag, 1970. 




\smallskip \noindent \hangindent 20 pt Wang, Ting, and Virginia R. Young. 2012a. {\it Optimal commutable annuities to minimize the probability of lifetime ruin.} {\it Insurance: Mathematics and Economics}, 50(1): 200-216.

\smallskip \noindent \hangindent 20 pt Wang, Ting, and Virginia R. Young. 2012b. {\it Maximizing the utility of consumption with commutable life annuities.} {\it Insurance: Mathematics and Economics}, 51(2): 352-369.

\smallskip \noindent \hangindent 20 pt Young, Virginia R. 2004. {\it Optimal Investment Strategy to Minimize the Probability of Lifetime Ruin.} {\it North American Actuarial Journal}, 8(4): 105-126.

\newpage

\sect{Appendix A: Auxiliary Functions}

The functions $g_0$, $g_1$, $h_0$, $h_1$, and $h_2$ in \eqref{eq:ODE1} are given as follows:
\[
\begin{split}
	g_0(z) &= (1-\alpha)\frac{c}{r}(z^{B_2}-z^{B_1})\Big[(B_2-1)z^{B_1-1}-(B_1-1)z^{B_2-1}\Big],\\
	\\
	g_1(z) &= (z^{B_2}-z^{B_1})\Big[B_1-B_2+\alpha(B_2-1)z^{B_1-1}\\
	&\hphantom{= (z^{B_2}-z^{B_1})\Big[B_1}-\alpha(B_1-1)z^{B_2-1}+\alpha(B_1-1)(B_2-1)(z^{B_1-1}-z^{B_2-1})\Big],\\
	\\
	h_0(z) &= (1-\alpha)^2\left(\frac{c}{r}\right)^2(B_1-B_2)z^{B_1+B_2-2}\Big[(B_1-1)z^{B_2-1}-(B_2-1)z^{B_1-1}\Big],\\
	\\
	h_1(z) &= (1-\alpha)\frac{c}{r}\bigg\{\Big[(B_2-1)z^{B_1-1}-(B_1-1)z^{B_2-1}\Big]\times\\
	&\hphantom{= (1-\alpha)\frac{c}{r}\bigg\{\Big[}\Big[(B_1-1)z^{B_1-1}-(B_2-1)z^{B_2-1}-\alpha(B_1-B_2)z^{B_1+B_2-2}\Big]\\
	&\hphantom{= (1-\alpha)\frac{c}{r}\bigg\{\Big[} - (B_1-B_2)z^{B_1+B_2-2}\Big[B_1-B_2+\alpha(B_2-1)z^{B_1-1}-\alpha(B_1-1)z^{B_2-1}\Big]\bigg\},\\
	\\
	\text{and }\hphantom{1em}&\\
	h_2(z) &= \Big[(B_1-1)z^{B_1-1}-(B_2-1)z^{B_2-1}-\alpha(B_1-B_2)z^{B_1+B_2-2}\Big]\times\\
	&\hspace{9em}\Big[B_1-B_2+\alpha(B_2-1)z^{B_1-1}-\alpha(B_1-1)z^{B_2-1}\Big].
\end{split}
\]

\setcounter{equation}{0}
\setcounter{section}{2}
\renewcommand{\theequation}{\Alph{section}.\arabic{equation}}
\sect{Appendix B: Proof of Proposition 5.4}


Consider $g_0$, $g_1$, $h_0$, $h_1$ and $h_2$ as in Appendix A and define $F$, $G$ and $H$ by
$$
	F(m,z) = {H(m,z) \over G(m,z)} := {h_2(z)(c/r-m)^2+h_1(z)(c/r-m)+h_0(z)\over g_1(z)(c/r-m) + g_0(z)},
$$
such that ODE \eqref{eq:ODE1} becomes $z'(m) = 1/F\big(m,z(m)\big)$. The following results hold for $F$, $G$, and $H$. We omit their elementary proofs.

\lem{B.1}{For $x(m)$ of Lemma 5.2, we have
$$
H\big(m,1/x(m)\big)=0;\quad \forall 0<m<c/r,
$$
and $H(m,z)>0$ (resp. $H(m,z)<0$) for $z>1/x(m)$ (resp. $z<1/x(m)$). 
\qed}

\lem{B.2}{ Assume $\xi(z):= g_0(z)/g_1(z) + c/r$. Then, we have
\item{$(i)$} $G\big(\xi(z),z\big)=0$, for $0<z<c/r$, and $G(m,z)>0$ (resp. $G(m,z)<0$) for $m<\xi(z)$ (resp. $m>\widehat{m}(z)$),
\item{$(ii)$} $\xi(z)\le \widehat{m}$, for $0<z<1$,
\item{$(iii)$} $\xi\big(1/x(\widehat{m})\big) = \widehat{m}$, i.e. the graphs of functions $z=1/x(m)$ and $m=\xi(z)$ intersect at $\big(\widehat{m},1/x(\widehat{m})\big)$; and,
\item{$(iv)$} $\xi(z) > x^{(-1)}(1/z)$ for $1/x(\widehat{m})<z<1/x(0)$.
\qed}

\lem{B.3}{$F(m,z)=H(m,z)/G(m,z)$ does not have a limit at points $(\widehat{m},1/x(\widehat{m}))$ and $(c/r,0)$.\qed}

%

\begin{figure}[t]
\centerline{
\adjustbox{trim={0.05\width} {0.0\height} {0.04\width} {0.06\height},clip}{\includegraphics[scale=0.35,page=1]{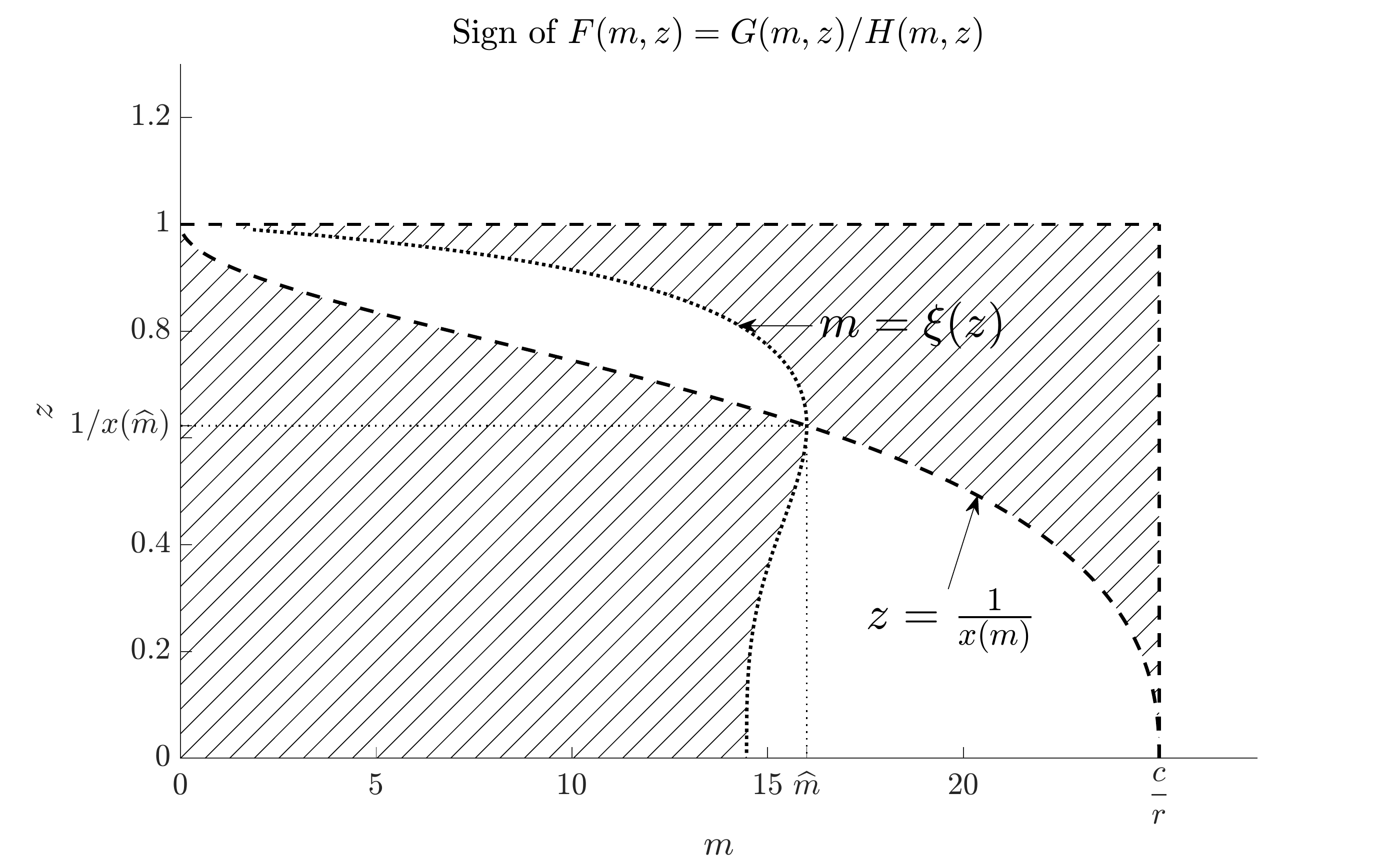}}
\hspace{0.25em}
\adjustbox{trim={0.05\width} {0.0\height} {0.04\width} {0.06\height},clip}{\includegraphics[scale=0.35,page=1]{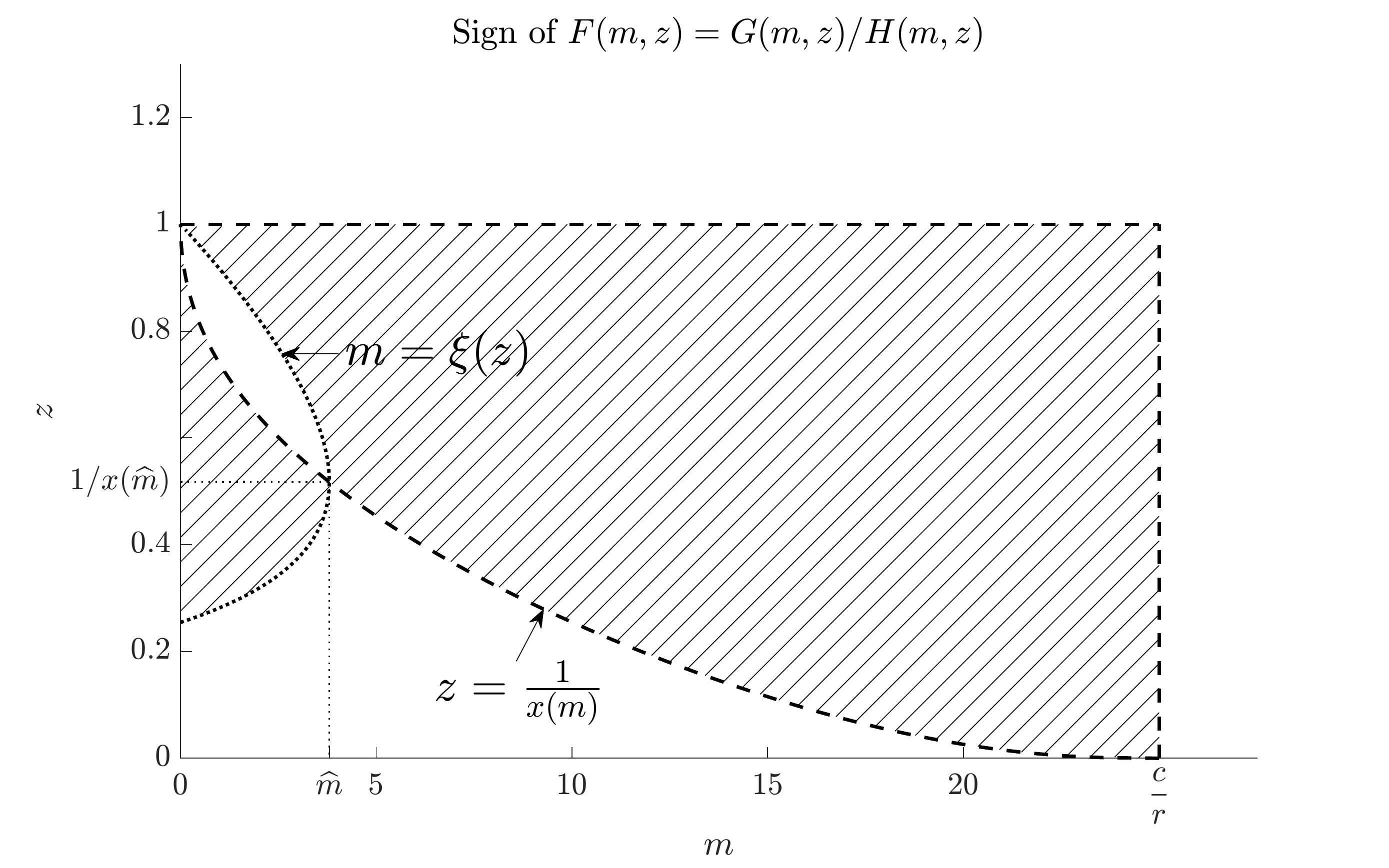}}
}
\nobreak
\centerline{{\bf Figure 9}: The sign of $F(m,z)=H(m,z)/G(m,z)$ in the rectangle $(m,z)\in[0,c/r]\times [0,1]$. The hatched }
\centerline{(resp. unhatched) region is where $F(m,z)<0$ (resp. $F(m,z)>0$).}
\end{figure}

Figure 6 illustrates the results of Lemmas B.1-3 for the numerical example of Section 5. Note, in particular, that at least one of the function $H(m,z)/G(m,z)$ or $G(m,z)/H(m,z)$ is continuous at any point $(m,z)\in \mathcal{D}_0\backslash \{(\widehat{m},1/x(\widehat{m})),(c/r,0)\}$. Thus, the classical theory for ordinary differential equations yields the following result concerning the existence and uniqueness of solutions of \eqref{eq:ODE1} in $\mathcal{D}_0$.

\lem{B.4} {For any $(m_0,z_0)\in\mathcal{D}_0\backslash\{(\widehat{m},1/x(\widehat{m})),(c/r,0)\}$, there exists a unique solution to ODE \eqref{eq:ODE1} in $\mathcal{D}_0$ satisfying $z(m_0)=z_0$. Furthermore, such solution extends on the left to the upper-left boundary $\overline{\partial \mathcal{D}}_0\cup\partial^-\mathcal{D}_0$ and on the right to the lower-right boundary $\underline{\partial\mathcal{D}}_0\cup\partial^+\mathcal{D}_0$.}

We may now prove Proposition 5.4. By Lemma B.4, for any $m_0\in (\underline{m}^*,\overline{m}^*)$, there exists a unique function $\tilde{z}(\cdot;m_0)$ satisfying ODE \eqref{eq:ODE1} in $\mathcal{D}_0$ and the initial condition $z(m_0)=1/x(m_0)$. By the comparison theorem for ordinary differential equations (Theorem V on page 65 of Walter (1970)), we must have $\underline{z}(m)<\tilde{z}(m;m_0)<\overline{z}(m)$ for values of $m$ where $\tilde{z}(\cdot;m_0)$, $\underline{z}$ and $\overline{z}$ are defined. It follows that $\tilde{z}(\cdot;m_0)$ must extend to the lower-right boundary $\underline{\partial\mathcal{D}}_0\cup\partial^+\mathcal{D}_0$. Now, define the constant $m^*$ by
$$
	m^* = \sup\Big\{m_0\in (\underline{m}^*,\overline{m}^*): \tilde{z}(\cdot;m_0)\text{ extends on the right to } \partial^+\mathcal{D}_0\Big\}.
$$
By the continuity of the solution of ODE \eqref{eq:ODE1} with respect to the initial data $z(m_0)=1/x(m_0)$, it easily follows that $\tilde{z}(c/r;m^*)=0$.

The uniqueness of the minimum probability of drawdown and Theorem 5.6 then yield that $\tilde{z}(\cdot;m^*)$ is the only solution of $\mathcal{D}_0$ satisfying the terminal condition $z(c/r) = 0$.

Finally, note that by Lemmas B.1 and B.2.(i), the right side of ODE \eqref{eq:ODE1} is negative in a neighborhood of $(m^*,1/x(m^*))$. Thus, the integral curve passing through the point $(m^*,1/x(m^*))$ spirals back toward the point $\big(\widehat{m},1/x(\widehat{m})\big)$, and $\tilde{z}(\cdot;m^*)$ is not defined for $m<m^*$.\qed

\setcounter{equation}{0}
\setcounter{section}{3}
\renewcommand{\theequation}{\Alph{section}.\arabic{equation}}

\newpage
\sect{Appendix C: Proof of Theorem 5.10}

It suffices to show that $\phi$ satisfies the conditions of Theorem 3.1. Because both $\Psi$ and $\Phi$ are probabilities of drawdown, as shown in Propositions 5.5 and 5.9, respectively, it then follows from Corollary 3.3 that $\phi$ is the {\it minimum} probability of drawdown on $\D$. 

Conditions of Theorem 3.1 have already been verified for $m^*\le m<c/r$ in the proof of Proposition 5.5. It remains to verify them for $0< m< m^*$.

First, we show that $\phi$ is continuous at $m=m^*$. From \eqref{eq:xm} and \eqref{eq:yhat_alm} it follows that:
\begin{equation}\label{eq:yalm2}
\eqalign{
& {1 \over \hat{y}_{\al m}(m)} \, {B_1 B_2 \over B_1 - B_2} \left( x(m)^{1-B_1} - x(m)^{1-B_2} \right)  \cr
& \quad = \left( {c \over r} - m \right) - \left( {c \over r} - \al m \right) \left[ {B_1 (1 - B_2) \over B_1 - B_2} x(m)^{1-B_1} + {B_2(B_1 - 1) \over B_1 - B_2} x(m)^{1-B_2} \right].}
\end{equation}
By the definition of $m^*$ in Proposition 5.4, we have $z(m^*)=1/x(m^*)$. Setting $m=m^*$ in \eqref{eq:yalm1} and \eqref{eq:yalm2} yields that $\tilde{y}_{\al m}(m^*)=\hat{y}_{\al m}(m^*)$ and, in turn, by \eqref{eq:phiTilde} and \eqref{eq:hatphi_sol} it follows that $\tilde{\phi}(y,m^*)=\hat{\phi}(y,m^*)$ for all $y\in[\tilde{y}_m(m^*),\tilde{y}_{\al m}(m^*)]=[\hat{y}_m(m^*),\hat{y}_{\al m}(m^*)]$. Thus, \eqref{eq:Psi} and \eqref{eq:Phi} yield $\Phi(w,m^*)=\Psi(w,m^*)$ for $\al\,m^*\le w\le m^*$, which implies that $\phi$ is continuous at $m=m^*$. 

Similarly, that $\tilde{\phi}(y,m^*)=\hat{\phi}(y,m^*)$ for all $y\in[\tilde{y}_m(m^*),\tilde{y}_{\al m}(m^*)]=[\hat{y}_m(m^*),\hat{y}_{\al m}(m^*)]$ yields that $\phi_{w}(w,m^*)$ and $\phi_{ww}(w,m^*)$ are continuous for $\al\,m^*\le w\le m^*$, except for the second derivative at $w=m^*$ where $\lim_{w\to m^{*-}}\phi_{ww}(w,m)=+\infty$. On the other hand, $\phi_m$ does not exist for $m=m^*$, since the right derivative of $z(m)$ and the derivative of $1/x(m)$ do not match at $m=m^*$ (see Figure 3). Nonetheless, for the right derivative, since $\Psi$ solves BVP \eqref{eq:BVP1}, we have $h^+_m(m^*,m^*) = \Psi^+_m(m^*,m^*)=0$; and, as we will show bellow, $h^-_m(m^*,m^*) = \Phi^-_m(m^*,m^*)>0$. Note that Theorem 3.1 allows for these irregularities. Namely, condition (i) allows for $\lim_{w\to m^-}\phi_{ww}(w,m)=+\infty$ and condition (ii) allows for non-differentiability of $\phi(w,\cdot)$ at $m=m^*$.

Next, we consider the case $0<m<m^*$ where $\phi=\Phi$. It is clear that $\Phi$ satisfies conditions (i) and (ii) of Theorem 3.1.  Also, from \eqref{eq:BVP2}, we see that $\Phi(\al m, m) = 1$, so condition (iv) holds.  $\Phi$ satisfies condition (vi) because it solves the HJB equation $\min_\pi {\cal L}^\pi \Phi = 0$.  Condition (v) is moot because $m < c/r$.

It only remains to show that $\Phi$ satisfies condition (iii) when $m \le m^*$.  Now, $\Phi_m(m, m) \ge 0$ if and only if $\hat \phi_m(\hat{y}_m(m), m) \ge 0$.  One can show that
$$
\hat \phi_m(\hat{y}_m(m), m) = {c - rm \over \la} \; \hat{y}_m'(m) + {\la - r \over \la} \; \hat{y}_m(m).
$$
After a long calculation, one obtains
$$
\hat{y}_m'(m) =  {\hat{y}_m(m) \over c/r - m} \left( 1 - \hat{y}_{\al m}(m) \, {c \over r} \, (1 - \al) \right),
$$
from which it follows that $\Phi_m(m, m) \ge 0$ if and only if
\begin{equation}\label{eq:yal_ineq}
\hat{y}_{\al m}(m) \le {\la \over c (1 - \al)}.
\end{equation}

We next show that inequality \eqref{eq:yal_ineq} holds exactly when $m \le \widehat{m}$, with $\widehat{m}\in(0,c/r)$ as in Lemma 5.2 (note that by Proposition 5.4, $m^*<\widehat{m}$). To this end, we begin by showing that $\hat{y}_{\al m}(m)$ strictly increases with respect to $m$ on $(0, c/r)$, with $\hat{y}_{\al m}(0+) \le {\la \over c (1 - \al)}$ and $\hat{y}_{\al m}((c/r)-) > {\la \over c (1 - \al)}$.  First, rewrite the expression for $\hat{y}_{\al m}(m)$ in \eqref{eq:yhat_alm} by substituting for $c/r - \al m$ from \eqref{eq:xm} and then simplifying to obtain
\begin{equation}\label{eq:Aux3}
{1 \over \hat{y}_{\al m}(m)} = \left( {c \over r} - m \right) {(B_1 - 1)(1 - B_2) \over B_1 - B_2} \left( {x(m)^{B_1 - 1} \over B_1} - {x(m)^{B_2 - 1} \over B_2} \right) .
\end{equation}
Differentiate \eqref{eq:Aux3} with respect to $m$ to get
\begin{equation}\label{eq:Aux4}
\eqalign{
\hat{y}'_{\al m}(m) \propto & \left( {x(m)^{B_1 - 1} \over B_1} - {x(m)^{B_2 - 1} \over B_2}  \right) \cr
& - \left( {c \over r} - m \right) \left( {B_1 - 1 \over B_1} x(m)^{B_1 - 1} + {1 - B_2 \over B_2} x(m)^{B_2 - 1} \right) {x'(m) \over x(m)},}
\end{equation}\eqref{eq:Aux4}
in which $\propto$ denotes {\it positively proportional to}.  We obtain $x'(m)$ by differentiating \eqref{eq:xm}.
$$
{x'(m) \over x(m)} = {B_1 - B_2 \over (B_1 - 1)(1 - B_2)} {1 - \al \over x(m)^{B_1 - 1} - x(m)^{B_2 - 1}} {c/r \over (c/r - m)^2}.
$$
Substitute this expression into \eqref{eq:Aux4}, eliminate $m$ via \eqref{eq:xm}, and simplify to get
\begin{equation}\label{eq:Aux5}
\hat{y}'_{\al m}(m) \propto \al \left( {B_1 - 1 \over B_1} x(m)^{B_1 - B_2} + {1 - B_2 \over B_2} \right) - {B_1 - B_2 \over B_1 B_2} \, x(m)^{B_1 - 1}.
\end{equation}
When $x = 1$, the right side of \eqref{eq:Aux5} equals $-{B_1 - B_2 \over B_1 B_2} \, (1 - \al) > 0$.  The derivative of the right side of \eqref{eq:Aux5} with respect to $x$ is also positive.  Thus, because $x(m) > 1$ for all $m \in (0, c/r)$, it follows that $\hat{y}'_{\al m}(m) > 0$ on this interval.  In other words, $\hat{y}_{\al m}(m)$ strictly increases with respect to $m$ on $(0, c/r)$.

Now, $x(0+) = 1$; thus,
$$
\hat{y}_{\al m}(0+) = - {r \over c} \, {B_1 B_2 \over (B_1 - 1)(1 - B_2)} = {\la \over c} \le {\la \over c(1 - \al)}.
$$
Also, $x((c/r)-) = \infty$, so $\hat{y}_{\al m}((c/r)-)$ is indeterminate.  By applying L'H\^opital's rule when we take the limit $m \to (c/r)-$, we obtain
$$
\hat{y}_{\al m}((c/r)-) = {r \over c(1 - \al)} \, {B_1 \over B_1 - 1} > {\la \over c(1 - \al)}.
$$
It follows that for any fixed $\al \in [0, 1)$, there exists a unique $\widehat{m} \in [0, c/r)$ such that
\begin{equation}\label{eq:Aux6}
\hat{y}_{\al m}(\widehat{m}) = {\la \over c(1 - \al)} \, .
\end{equation}

We are left with showing that the $\widehat{m}$ that solves \eqref{eq:Aux6} also solves \eqref{eq:mhat}.  Substitute the expression for $\hat{y}_{\al m}(\widehat{m})$ in \eqref{eq:Aux6} into \eqref{eq:Aux3}, solve \eqref{eq:xm} for $x(\widehat{m})^{B_2 - 1}$, substitute that expression into \eqref{eq:Aux3}, solve the result for $x(\widehat{m})^{B_1 - 1}$ to obtain
\begin{equation}\label{eq:Aux7}
x(\widehat{m})^{B_1 - 1} = \al B_1 + {{c \over r}(1 - \al) \over {c \over r} - \widehat{m}}.
\end{equation}
By symmetry, the analogous expression for $x(\widehat{m})^{B_2 - 1}$ also holds, namely with $B_1$ replaced by $B_2$, and \eqref{eq:mhat} follows when we equate these two expressions for $x(\widehat{m})$.\qed

\bye